\documentclass{article}

\usepackage{bm}
\usepackage{amsmath}
\usepackage{tabularx}
\usepackage{array}
\usepackage{booktabs}
\usepackage{multirow}
\usepackage{float}
\usepackage{graphicx}
\usepackage{amsfonts}
\usepackage{color}
\usepackage{url}
\usepackage{algorithm}
\usepackage{algorithmic}
\usepackage{enumitem}
\usepackage{comment}
\usepackage{ulem}
\usepackage{authblk}
\usepackage{subfigure}
\usepackage{tablefootnote}
\usepackage[numbers,sort&compress]{natbib}

\newcolumntype{L}[1]{>{\raggedright\let\newline\\\arraybackslash\hspace{0pt}}m{#1}}
\newcolumntype{C}[1]{>{\centering\let\newline  \\\arraybackslash\hspace{0pt}}m{#1}}
\newcolumntype{R}[1]{>{\raggedleft\let\newline \\\arraybackslash\hspace{0pt}}m{#1}}

\title{Emerging Drug Interaction Prediction
	Enabled by Flow-based Graph Neural Network with Biomedical Network}

\author[a]{Yongqi Zhang}

\author[b]{Quanming Yao \thanks{Corresponding author: qyaoaa@tsinghua.edu.cn}}

\author[b]{Ling Yue}

\author[c]{Xian Wu}

\author[c]{Ziheng Zhang}

\author[c]{Zhenxi Lin}

\author[c]{Yefeng Zheng}

\affil[a]{4Paradigm Inc., Beijing, China}

\affil[b]{Department of Electronic Engineering, Tsinghua University, Beijing, China}

\affil[c]{Tencent Jarvis Lab, Shenzhen, China}

\date{}
\begin{document}

\maketitle

\begin{abstract}
Accurately predicting drug-drug interactions (DDI) for emerging drugs, which offer possibilities for treating and alleviating diseases, with computational methods can improve patient care and contribute to efficient drug development. However, many existing computational methods require large amounts of known DDI information, which is scarce for emerging drugs. In this paper, we propose EmerGNN, a graph neural network (GNN) that can effectively predict interactions for emerging drugs by leveraging the rich information in biomedical networks. EmerGNN learns pairwise representations of drugs by extracting the paths between drug pairs, propagating information from one drug to the other, and incorporating the relevant biomedical concepts on the paths. The different edges on the biomedical network are weighted to indicate the relevance for the target DDI prediction. Overall, EmerGNN has higher accuracy than existing approaches in predicting interactions for emerging drugs and can identify the most relevant information on the biomedical network.
\end{abstract}

\newpage


\section{Introduction}
\label{sec:intro}

Science advancements and regulatory changes have led to the development of numerous emerging drugs worldwide, particularly for rare, severe, or life-threatening illnesses~\cite{su2022trends,ledford2022hundreds}. 
These drugs are novel substances with unknown or unpredictable risks, as they have not been extensively regulated or used before.
For instance, although hundreds of COVID-19 drugs have been developed, only six have been recommended by the FDA as of Oct 2023, 
such as dexamethasone and hydrocortisone. 
Clinical deployment of new drugs is cautious and slow, making it crucial to identify drug-drug interactions (DDIs) for emerging drugs. 
To speed up the discovery of potential DDIs, computational techniques, particularly machine learning approaches, have been developed~\cite{percha2013informatics,vilar2014similarity,tanvir2021predicting,yu2021sumgnn}.
However, with limited clinical trial information, unexpected polypharmacy or side effects can be severe and difficult to detect~\cite{Letinier2019,Jiang2022}.

	Early DDI prediction methods used fingerprints~\cite{rogers2010extended}
	or hand-designed features \citep{vilar2014similarity,dewulf2021cold}
	to indicate interactions based on drug properties.
	Although these methods can work directly on emerging drugs
	in a cold-start setting \citep{liu2022predict,dewulf2021cold},
	they can lack expressiveness and ignore the mutual information between drugs.
	DDI facts can naturally be represented as a graph where nodes represent drugs
	and edges represent interactions between a pair of drugs. 
	Graph learning methods can learn 
	drug embeddings for prediction \citep{yao2022effective},
	but they rely on historical interactions,
	thus cannot address the problem of scarce interaction data for emerging drugs.

	Incorporating large biomedical networks as side information for DDI prediction is an alternative to learning solely from DDI interactions \citep{zitnik2018modeling,karim2019drug,tanvir2021predicting,huang2020skipgnn,yu2021sumgnn,lin2020kgnn,ren2022biomedical}.
	These biomedical networks, 
	such as HetioNet \citep{himmelstein2017systematic},
	organize facts into a directed multi-relational graph,
	recording relationships between biomedical concepts,
	such as genes, 
	diseases, and drugs.
	Tanvir et. al. used hand-designed meta-paths from the biomedical network~\cite{tanvir2021predicting},
	while Karim et. al. learned embeddings from the network
	and used a deep network to do DDI prediction~\cite{karim2019drug}.
	Graph neural networks~\cite{kipf2016semi,gilmer2017neural} can obtain expressive node embeddings 
	by aggregating topological structure and drug embeddings,
	but existing methods~\citep{zitnik2018modeling,huang2020skipgnn,yu2021sumgnn,lin2020kgnn,ren2022biomedical} do not specially consider emerging drugs, leading to poor performance in predicting DDIs for them.

Here,
we propose to 
use large biomedical network
to predict DDI for emerging drugs
by learning from the biomedical concepts connecting target drugs pairs.
Although emerging drugs may not have sufficient interactions in the DDI network,
they often share the same 
biochemical concepts used in the drug development with existing drugs,
such as targeted genes or diseases.
Therefore,
we exploit related paths from
the biomedical networks for given drug pairs. 
However, 
properly utilizing these networks can be challenging
as they are not developed for emerging drugs,
and the mismatch of objectives can lead machine learning models 
to learn distracting knowledge.

To accurately and interpretable predict DDI for emerging drugs,
we introduce EmerGNN, a GNN method
that learns pair-wise drug representations
by integrating the biomedical entities and relations connecting them. 
A flow-based GNN architecture
extracts paths connecting drug pairs,
traces from an emerging drug to an existing drug,
and integrates information of the biomedical concepts
along the paths.
This approach utilizes
shared information in both biomedical and interaction networks.
To extract relevant information,
we weight different types of relations on the biomedical network,
and edges with larger weights on the paths are helpful for interpretation.
Compared with other GNN-based methods,
EmerGNN propagates on the local subgraph around the drug pair to be predicted
and better discovers directional information flow within the biomedical network.
In summary, our main contributions are as follows:
\begin{itemize}
	\item 
		Building upon a biomedical network,
		we develop an effective deep learning method
		that predicts interactions for emerging drugs accurately.	
	
	\item We propose  EmerGNN, 
	a GNN-based method that learns 
	pair-wise representations of drug pairs 
	to predict DDIs for emerging drugs
	by
	integrating the relevant biomedical concepts connecting them.
	
	\item Extensive experiments show that EmerGNN is effective in
	predicting interactions for emerging drugs.
	The learned concepts on the biomedical network
	are interpretable.
	
	\item 
	\textcolor{black}{EmerGNN's strong prediction ability has the potential to clinically 
		improve patient care and contribute to more efficient drug development processes.}
\end{itemize}

\section{Results}
\label{sec:results}

\begin{figure}[H]
	\centering
	\includegraphics[width=0.95\textwidth]{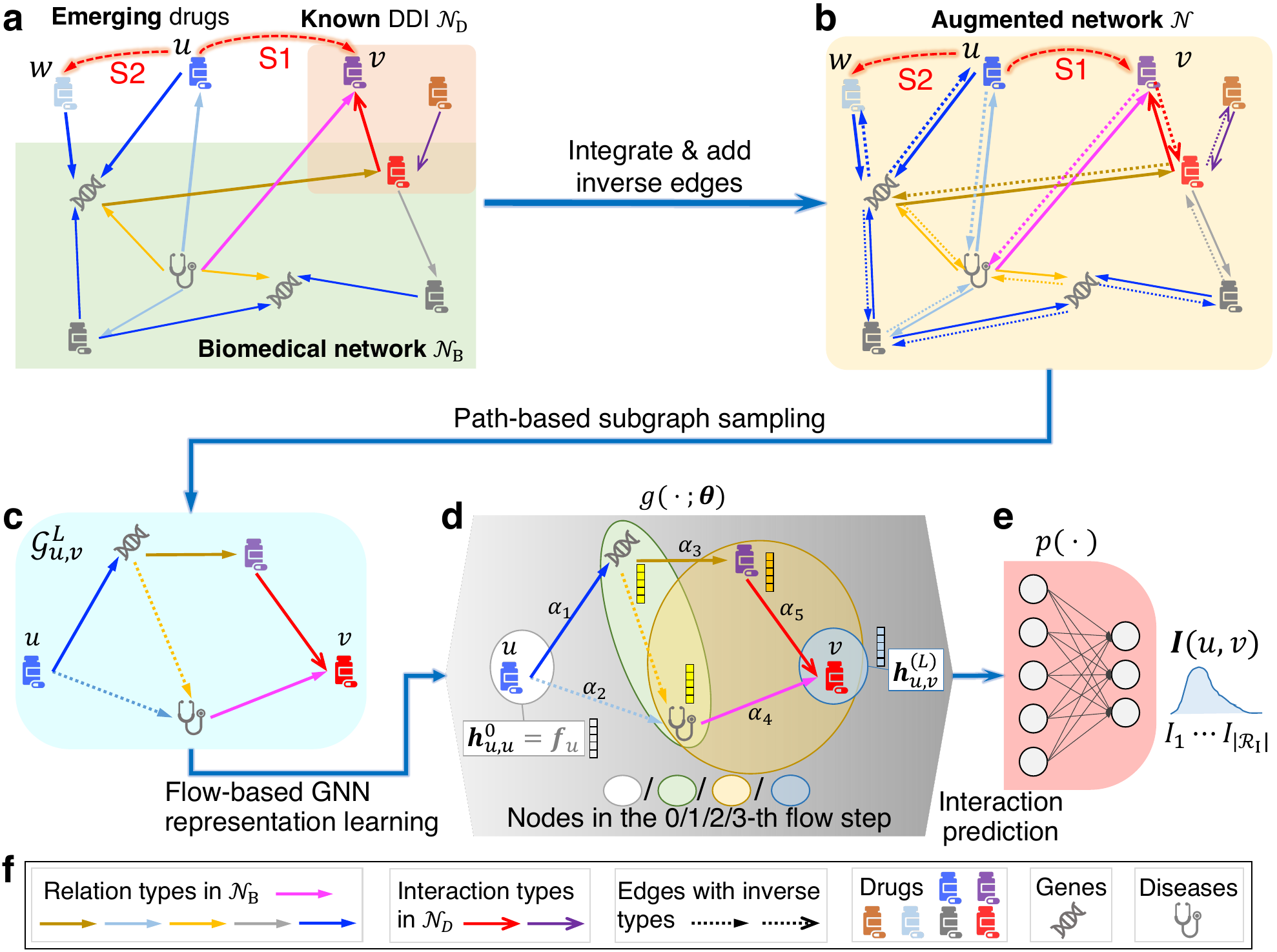}
	\caption{
		\textbf{
		Overview of EmerGNN.}
		\textbf{(a)} 
		Problem formulation:
		Given a DDI network $\mathcal N_\text{D}$ of existing drugs
		and a large biomedical network $\mathcal N_\text{B}$
		providing side information for the drugs,
		the task is to predict the interaction type 
		between 
		an emerging drug (like $u$ in dark blue) and an existing drug (like $v$ in purple)
		in the S1 setting,
		or interaction type
		between two emerging drugs
		(like  $u$ in dark blue and $w$ in light blue)
		in the S2 setting.
		\textbf{(b)} 
		Augmented network $\mathcal N$:
		The DDI network and biomedical network are integrated and 
		edges with inverse types are incorporated
		to obtain an augmented network $\mathcal N$.
		The augmentation brings better communication
		among drugs and entities in both interaction and biomedical networks.
		\textbf{(c)}
		Path-based subgraph:
		Given a drug pair $(u,v)$ to be predicted,
		all the paths from $u$ to $v$ with length no larger than $L$ 
		are extracted to construct a path-based subgraph $\mathcal{G}_{u,v}^L$.
		\textbf{(d)}
		Flow-based GNN $g(\cdot;\bm \theta)$ with parameters $\bm \theta$:
		the network flows the initial drug features $\bm h_{u,u}^0=\bm f_u$
 		over essential information in $\mathcal G_{u,v}^L$ for $L$ steps.
 		It uses different attention weights $\alpha$'s to weight the importance of different edges.
 		After $L$ steps, a pair-wise representation $\bm h_{u,v}^L$ between $u$ and $v$ is obtained as the subgraph representation of $\mathcal{G}_{u,v}^L$. 
 		\textbf{(e)}
 		Interaction predictor $p(\cdot)$:
 		a simple linear classifier $p(\bm h_{u,v}^L)$ 
 		outputs a distribution $\bm I(u,v)$,
 		where each dimension indicates
 		an interaction type $i\in\mathcal R_\text{I}$ between $u$ and $v$.
 		\textbf{(f)}
 		Legends:
 		The different relation and interaction types are indicated by arrows with different colors. Edges with inverse types are indicated by dashed arrows with corresponding color.
 		The icons represent biomedical concepts including drugs, genes and diseases.
 	}
	\label{fig:pipeline}
\end{figure}

\paragraph{EmerGNN: encoding pair-wise representations with flow-based GNN for emerging drugs.}
We focus on two DDI prediction task settings for emerging drugs
\cite{dewulf2021cold,liu2022predict,yu2022stnn} (Fig.~\ref{fig:pipeline}a, Method):
S1 setting,
determining the interaction type between an emerging drug and an existing drug,
and S2 setting,
determining the interaction type between two emerging drugs.
To connect emerging and existing drugs,
we use a large biomedical network HetioNet \cite{himmelstein2017systematic},
which contains entities and relations related to biomedical concepts.
We assume that all the emerging drugs are connected to entities
in the biomedical network,
allowing us to infer their properties from existing drugs 
and the biomedical network.

Given the DDI network and biomedical network (Fig.~\ref{fig:pipeline}a),
we firstly integrate the two networks to enable communication
between
existing and emerging drugs connected by biomedical concepts, 
such as
proteins, diseases or other drugs,
and then add inverse edges by introducing inverse types for 
each relation and interaction type.
The two steps generate an augmented network where
the drugs and biomedical entities can communicate better (Fig.~\ref{fig:pipeline}b).
For a target drug pair to be predicted
(for example an emerging drug $u$ and an existing drug $v$),
we extract all the paths with length no longer than 
$L$ between them,
and combine the paths to form a 
path-based subgraph $\mathcal{G}_{u,v}^L$ (Fig.~\ref{fig:pipeline}c).
The value of $L$ is a hyper-parameter to be tuned 
(Supplementary Table~2).
A flow-based 
GNN $g(\cdot;\bm\theta)$ with parameters $\bm \theta$ (Fig.~\ref{fig:pipeline}d)
is applied on $\mathcal{G}_{u,v}^L$ 
to trace drug features $\bm h_{u,u}^0=\bm f_u$ (like fingerprints)
along the biomedical edges and integrate essential information along the path.
In each iteration $\ell$,  the GNN flows to drug-specific entities
that are $\ell$-steps away from drug $u$ and $(L-\ell)$-steps away from drug $v$ in the augmented network.
An attention mechanism is applied on the edges in $\mathcal{G}_{u,v}^L$ 
to adjust their importance.
The GNN iterates $L$ steps
to return the pair-wise representation $\bm h_{u,v}^{(L)}$.
Finally,
$\bm h_{u,v}^{(L)}$ is fed to a linear classifier $p(\cdot)$
to predict the interaction type between $u$ and $v$ (Fig.~\ref{fig:pipeline}e).

\paragraph{Comparison of EmerGNN to baseline methods in DDI prediction.}
Two public datasets DrugBank~\citep{wishart2018drugbank}
and TWOSIDES~\citep{tatonetti2012data}
are used.
The original drug set is split into three parts 
with a ratio of $7$:$1$:$2$
for training, validation, and testing (Method).
The drugs in validation and testing sets are 
considered emerging drugs for validation and testing, respectively.
For the DrugBank dataset,
there is at most one interaction type between any drug pair,
and the task is to predict the exact type 
in a multi-type classification setting.
Macro F1-score,
Accuracy, 
and Cohen's Kappa \citep{cohen1960coefficient} are used
as performance metrics,
with F1-score as the primary metric.
For TWOSIDES dataset,
there may be multiple interaction types between a drug pair,
and the task is to predict whether a pair of drugs will have a certain interaction type
under a binary classification setting.
PR-AUC, ROC-AUC and Accuracy are used to evaluate the performance,
with PR-AUC being the primary metric.

In S1 setting,
\textit{Emb} type methods, particularly \texttt{MSTE},
poorly predict emerging drugs because their embeddings are not updated during training.
\texttt{KG-DDI} performs better as it updates drug embeddings with information in the biomedical network.
For \textit{DF} methods, \texttt{CSMDDI} and \texttt{STNN-DDI} outperform MLP
on {DrugBank} dataset
with their designed training schemes in a cold-start setting,
but they do not perform well on TWOSIDES with more interaction types.
\texttt{HIN-DDI} outperforms \texttt{MLP},
indicating that graph features from biomedical network can benefit DDI prediction.
Deep \textit{GNN}-based methods 
may not perform better than \textit{DF} methods on DrugBank
since the GNN-based methods may not well capture
the crucial property of similarity for emerging drug prediction (Fig.~\ref{fig:analyzing}).
\texttt{CompGCN}, \texttt{Decagon} and \texttt{KGNN} 
perform comparably due to their similar GNN architecture design.
\texttt{SumGNN} constrains message passing in the enclosing subgraph between drug pairs,
making information more focused.
\texttt{DeepLGF} is the best GNN-based baseline
	by fusing information from multiple sources,
	taking the advantage of both drug features and graph features.
\texttt{EmerGNN} 
significantly outperforms all compared methods
as indicated by the small p-values under two-sided t-testing of statistical significance.
First,
by learning paths between emerging and existing drugs,
it can capture the graph features,
whose importance has been verified by the \textit{GF} method \texttt{HIN-DDI}.
Second,
different from \texttt{CompGCN}, \texttt{Decagon}, \texttt{KGNN},
and \texttt{DeepLGF},
the importance of edges can be weighted such that it can implicitly learn the similarity properties (Fig.~\ref{fig:analyzing}).
Third,
with the designed path-based subgraph
and flow-based GNN architecture,
EmerGNN captures more relevant information from the biomedical network,
thus outperforming \texttt{CompGCN} and \texttt{SumGNN} (Supplementary Figure~4) as well.

We evaluate the top-performing models in each type
in the more challenging S2 setting (Table~\ref{tab:emerging}),
where both drugs are new with sparser information.
While \texttt{KG-DDI} and \texttt{DeepLGF}
performed well in S1 setting,
they struggled in S2 setting since they need to learn representations of both drugs effectively.
Conversely,
\texttt{CSMDDI} and \texttt{HIN-DDI} 
performed more consistently,
with \texttt{CSMDDI} ranking second on DrugBank 
and \texttt{HIN-DDI} ranking second on TWOSIDES.
This may be due to their simple models but effective features.
In comparison,
\texttt{EmerGNN} significantly outperforms all the baselines
under two-sided t-testing of statistical significance
by aggregating essential information from the biomedical network.
Additionally,
we provide results for the S0 setting (Supplementary Table~3),
which predicts interactions between existing drugs.
We thoroughly investigate why \texttt{EmerGNN} has superior performance for DDI prediction in the following results.

\begin{figure}[H]
	\centering
	\color{black}
	\vspace{-8px}
%
	\includegraphics[width=\textwidth]{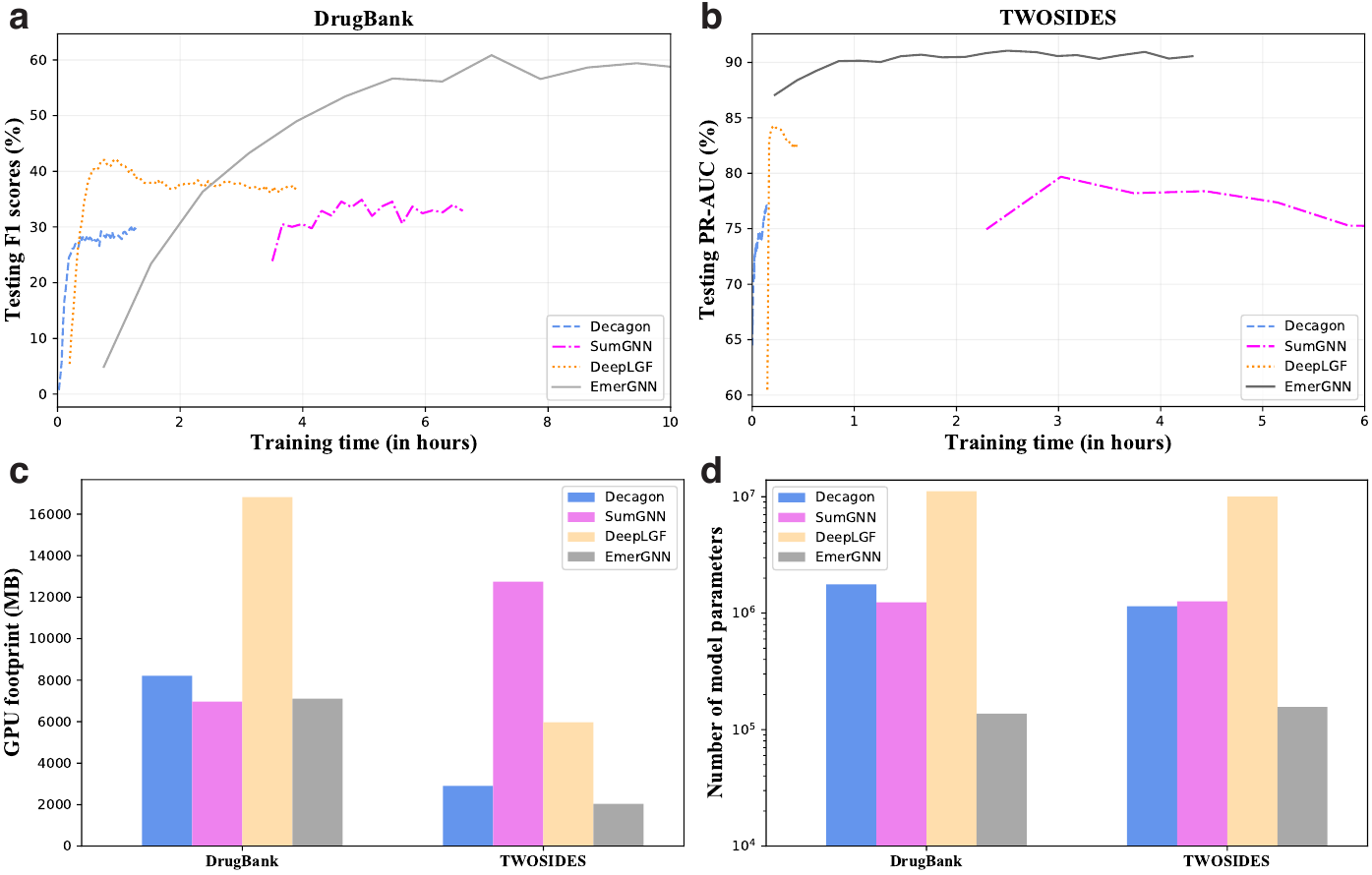}
	\caption{Complexity analysis of different GNN-based methods in the S1 setting.
		\textbf{(a)} Comparison of training curves on DrugBank dataset. 
		\textbf{(b)} Comparison of training curves on TWOSIDES dataset.
		\textbf{(c)}  Comparison of GPU memory footprint usage on the two datasets
		in MB.
		\textbf{(d)} Comparison of number of trainable model parameters on the two datasets.}
	\label{fig:complexity}
\end{figure}

\begin{figure}[H]
	\centering
	\includegraphics[width=\textwidth]{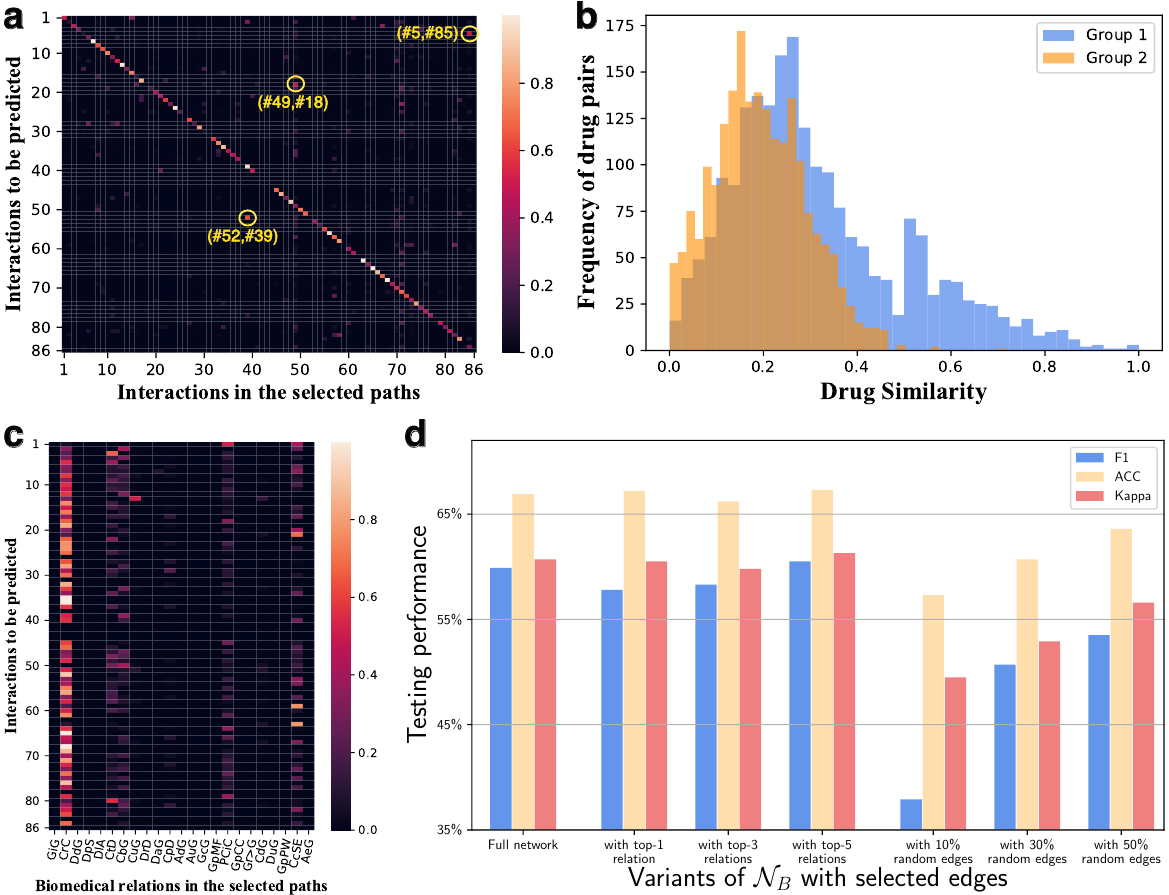}
	\caption{
		Analysis of relation types on the selected paths on the DrugBank dataset.
		In detail,
		we first extract top five paths from $u$ to $v$ and from $v$ to $u$, respectively,
		for each triplet $(u, i_\text{pred}, v)$ in testing
		based on the average attention weights of edges in each path
		with a beam search algorithm (Algorithm 2 in Supplementary Section 1).
		Next,
		we count how many times an interaction type
		$i\in\mathcal R_\text{I}$ and 
		a relation type $r\in\mathcal R_\text{B}$ appears on the selected edges given $i_\text{pred}$
		(Supplementary Figure~3).
		\textbf{(a)} Heatmap of correlation between interaction $i_\text{pred}$ to be predicted
		and interaction types $i$ in the selected paths.
		Yellow circles indicate the three highlighted interaction pairs outside  the diagonal (Supplementary Table~5).
		\textbf{(b)} 
		The histogram distribution of fingerprint similarities in \textit{Group 1} (a drug $u$ with another drug $u_1$, which connected to $v$ with interaction type $i_\text{pred}$)
		and \textit{Group 2} (a drug $u$ with a random drug $u_2$).
		\textbf{(c)} Heatmap of correlation between  interaction $i_\text{pred}$ to be predicted
		and biomedical relations $r$ in the selected paths.
		\textbf{(d)} Performance of modified biomedical networks with selected relations.
		Leftmost is the performance of EmerGNN with full biomedical network.
		The middle three parts are EmerGNN with 
		top-1 (CrC, with 0.4\% edges),
		top-3 (CrC, CbG, CsSE, with 9.3\% edges),
		top-5 (CrC, CtD, CvG, PCiC, CsSE, with 9.4\% edges) attended relations in the biomedical network.
		The right three parts are EmerGNN with randomly sampled 10\%, 30\%, 50\% edges from the biomedical network.
	}
	\label{fig:analyzing}
	\vspace{-5px}
\end{figure}

\paragraph{Analysis of the Computational complexity.}
Since \texttt{EmerGNN} learns pair-wise 
representations for each drug pair,
the computation complexity is higher than the other GNN-based methods.
However, \texttt{EmerGNN} can achieve higher accuracy than other GNN-based methods in just a few hours, 
and longer training time has the potential to achieve
even better performance (Fig.~\ref{fig:complexity}a-b).
Among the baseline GNN methods, 
\texttt{Decagon} is the most efficient
as it only uses information related to drug, protein and disease in the biomedical network.
\texttt{SumGNN} and  \texttt{EmerGNN} are slower than \texttt{Decagon} and  \texttt{DeepLGF}
as they need to learn specific subgraph representations 
for different drug pairs.
Given that the clinical development of a typical innovative drug usually takes years \citep{brown2021clinical}, 
the computation time of  \texttt{EmerGNN} is acceptable.
We also compare the GPU memory footprint (Fig.~\ref{fig:complexity}c) and
 the number of parameters (Fig.~\ref{fig:complexity}d) of these GNN-based models.
It is clear that  \texttt{EmerGNN} is memory and parameter efficient.
First,
its subgraphs for DDI prediction are much smaller than the biomedical network (Supplementary Figure~1).
Second,
\texttt{EmerGNN} mainly relies on the biomedical concepts instead of the drugs' embeddings to do predictions, 
resulting in a small number of parameters.
In comparison,
\texttt{DeepLGF} requires a large number of model parameters to learn embeddings from the biomedical network.

\paragraph{Analysis of drug interaction types in the learned subgraph.}
\texttt{EmerGNN} uses attention weights to measure the importance of edges in the subgraph 
for predicting DDI of the emerging drugs.
Here,
we analyze what is captured by the attention weights
by checking correlations between predicted
interaction types
with interactions and relations in the path-based subgraphs (Fig.~\ref{fig:analyzing}).

We firstly analyze the correlations 
between the interaction type $i_\text{pred}$ to be predicted
and  interaction types obtained in the selected paths.
The dominant diagonal elements in the heatmap (Fig.~\ref{fig:analyzing}a)
suggests that 
when predicting a target  interaction $i_\text{pred}$ for $(u,v)$, 
paths with larger attention weights in the subgraph $\mathcal{G}_{u,v}^L$ 
are likely to go through
another drug (for instance $u_1$) that has interaction $i_1 = i_\text{pred}$
with the existing drug $v$.
We suppose that these drugs like $u_1$ may have similar properties as the emerging drug $u$.
To demonstrate this point,
we group these cases of drug pairs $(u, u_1)$ as 
\textit{Group 1}
and other pairs $(u, u_2)$ with a random drug $u_2$ as \textit{Group 2}.
The distributions of drug fingerprints similarities show
that \textit{Group 1} has a larger quantity of 
highly similar drug pairs ($>0.5$) 
than \textit{Group 2} (Fig.~\ref{fig:analyzing}b),
demonstrating the crucial role of
similar drugs in predicting DDIs for emerging drugs,
and our method can implicitly search for these drugs.
Apart from the diagonal part, 
there exists strongly correlated pairs of interactions.
This happens, 
for example, 
when 
the emerging drug $u$ finds some connections with another drug $u_3$ whose intersection $i_3$ with the existing drug $v$ is correlated with $i_\text{pred}$.
In these cases,
we find strongly correlated pairs
like ``increasing constipating activity'' and ``decreasing analgesic activity'' (Fig.~\ref{fig:analyzing}a, Supplementary Table~5),
verified by Liu et al. \cite{liu2002low}.

We then analyze the biomedical relation types in the selected paths
by visualizing
correlations between
the interaction to be predicted $i_\text{pred}$ and biomedical relation types in the selected paths.
There are a few relation types
consistently selected when predicting different interaction types
(Fig.~\ref{fig:analyzing}c).
In particular,
the most frequent relation type is the drug resembling relation CrC,
which again verifies the importance of similar drugs for emerging drug prediction.
Other frequently selected types are related to diseases (CrD), genes (CbG),
pharmacologic classes (PCiC) and side effects (CsSE).
To analyze their importance,
we compare the performance of \texttt{EmerGNN}
with the full 
biomedical network,
and networks with only
top-1,
top-3,
or
top-5 attended relations
(the middle part of Fig.~\ref{fig:analyzing}d).
As a comparison,
we randomly sample 10\%, 30\% and 50\% edges from 
the biomedical performance
and show the performance 
(the right part of Fig.~\ref{fig:analyzing}d).
Keeping the top-1, top-3 and top-5 relations in 
biomedical network
can all lead to comparable performance as using a full network.
However,
the performance substantially deteriorates when edges
are randomly dropped.
These experiments show that
EmerGNN selects important and relevant relations in the biomedical network for DDI prediction.

\paragraph{Case study on drug-pairs.}

\begin{figure}[H]
	\centering
	\includegraphics[width=\textwidth]{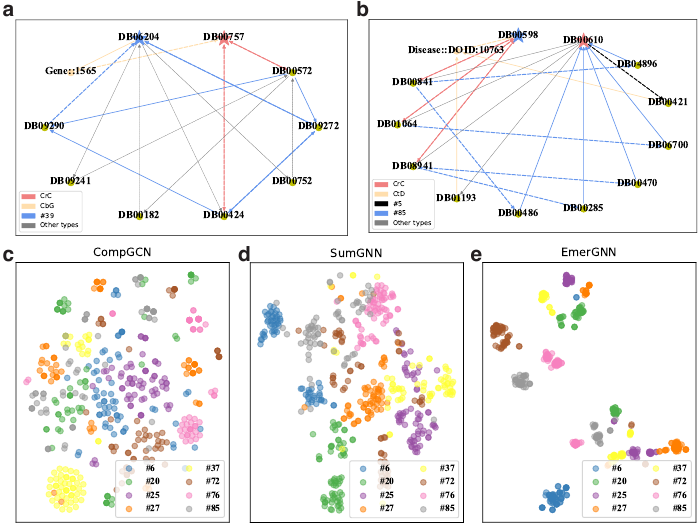}
	\caption{Visualization of drug pairs. 
		\textbf{(a-b)} 
		Two cases of subgraphs containing top ten paths according to the average of edges' attention weights on each path (explanations in Supplementary Table~6).
		The drug pairs to be predicted are highlighted as stars; 
		dashed lines mean reverse types; 
		CrC, CbG, CtD are biomedical relations;
		\#39, \#5, \#85 are interaction types;
		``other types'' in gray edges mean the interaction types aside from the given ones.
		In the first case,
		DB06204 (Tapentadol) in blue star is an existing drug ,
		and DB00757 (Dolasetron) in red star is an emerging drug.
		The target interaction type is ``\#Drug1 may decrease the analgesic activities of \#Drug2'' (\#52).
		In the second case,
		DB00598 (Labetalol) in blue star is an emerging drug,
		and DB00610 (Metaraminol) in red star is an existing drug.
		The target interaction type is ``\#Drug1 may decrease the vasoconstricting activities of \#Drug2'' (\#5).
		\textbf{(c-e)} t-SNE visualization~\cite{van2008visualizing}
		of the representations learned for drug pairs.
		As {CompGCN} embeds each entity separately,
		we concatenate embeddings of the two drugs' representations for a given drug pair.
		{SumGNN} encodes the enclosing subgraphs of $(u,v)$ for interaction prediction,
		thus we take the representation of enclosing subgraph
		as the drug pair representation.
		The drug pair representation of EmerGNN is directly given by 
		$\bm h_{u,v}^{(L)}$.
		Since there are too many interaction types and drug pairs in 
		$\mathcal N_\text{D-test}$,
		eight interaction types and sixty-four drug pairs are randomly sampled for each interaction type.
		The legends in these figures specify the IDs of the interaction type to be predicted;
		each dot denotes a DDI sample $(u, i, v)$;
		the different colors in dots indicate the interaction type $i$
		that the drug pairs $(u,v)$ have.}
\label{fig:visualization}
\end{figure}

We present cases of selected paths from subgraphs
by selecting top ten paths between $u$ and $v$
based on the average of edges' attention weights on each path (Fig.~\ref{fig:visualization}a-b).
In the first case,
there are interpretable paths supporting the target prediction
(Supplementary Table~6).
For example, 
there are paths connecting
the two drugs through the 
binding protein Gene::1565 (CYP2D6),
which is a P450 enzyme that plays a key role in drug metabolism \citep{estabrook2003passion}.
Another path finds a similar drug DB00424 (Hyoscyamine) of 
DB00757 (Dolasetron) 
through the resemble relation (CrC), 
and concludes that DB06204 (Tapentadol) 
may potentially decrease the analgesic activity of DB00757 (Dolasetron) 
due to the correlation between constipating and analgesic activities
(Fig.~\ref{fig:analyzing}a).
In the second case, 
we make similar observations (Supplementary Table~6).
In particular,
a path finds a similar drug DB00421 (Spironolactone) of DB00598 (Labetalol),
which may decrease the vasoconstricting activity of DB00610 (Metaraminol),
providing a hint that Labetalol may also decrease the vasoconstricting activity of Metaraminol.
Compared
with the original subgraphs $\mathcal G_{u,v}^L$ which have 
tens of thousands of edges (Supplementary Figure~1),
the learned subgraphs are much smaller
and more relevant to the target prediction.
More examples with detailed interpretations on the paths
can support that EmerGNN finds important paths
that indicate relevant interaction types and biomedical entities 
for emerging drug prediction (Supplementary Figure~5).

Next,
we visualize the drug pair representations obtained by 
\texttt{CompGCN},
\texttt{SumGNN} 
and \texttt{EmerGNN} (Fig.~\ref{fig:visualization}c-e).
As shown,
the drug pairs with the same interaction are more densely gathered in \texttt{EmerGNN} than \texttt{CompGCN} and \texttt{SumGNN}.
This means that the drug pair representations of \texttt{EmerGNN}
can better separate the different interaction types.
As a result,
\texttt{EmerGNN} is able to learn better representations than the other GNN methods, like \texttt{CompGCN} and \texttt{SumGNN}.

\paragraph{Ablation studies.}

\begin{figure}[H]
	\centering
\includegraphics[width=\textwidth]{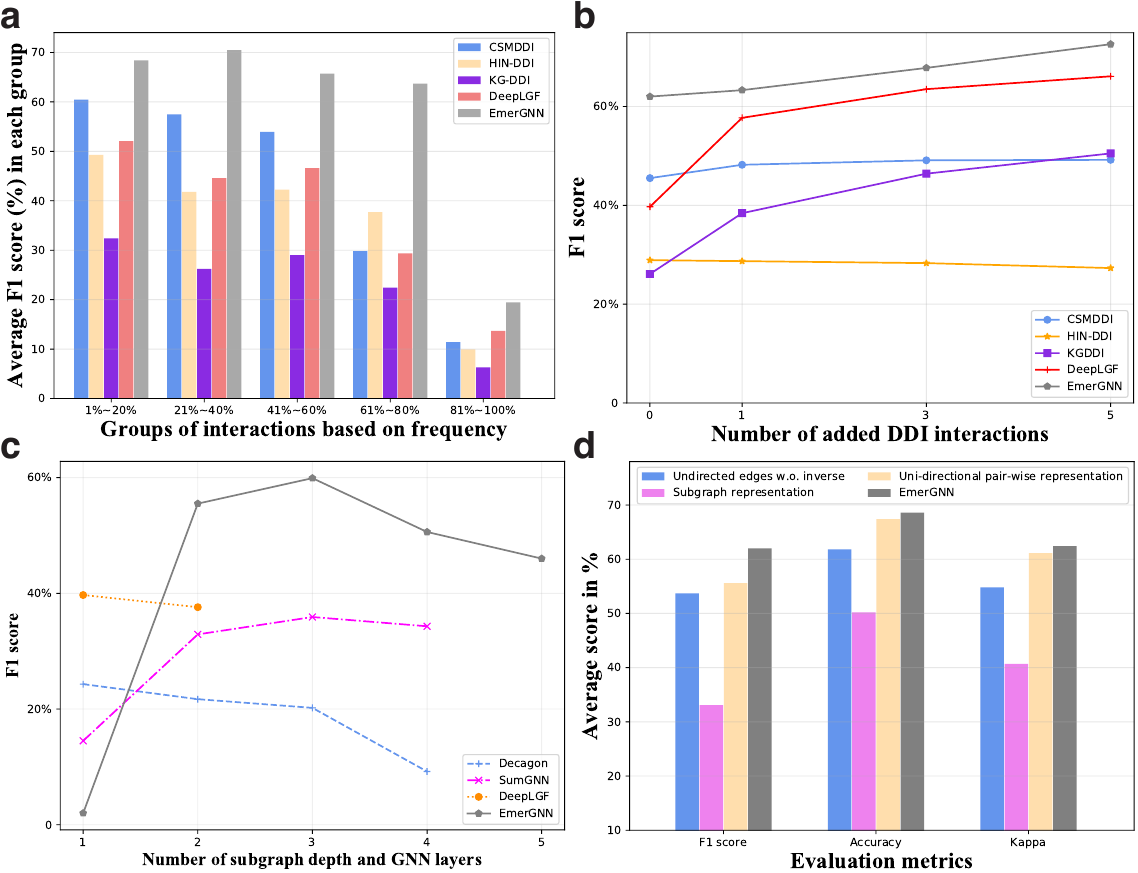}
\caption{Ablation studies on the DrugBank dataset. 
	\textbf{(a)} Performance comparison of interaction groups based on interaction frequency.
	The five groups are formed by 
	grouping the interaction types
	based on their frequency in the dataset,
	and the average macro F1 performance is shown in each group.
	\textbf{(b)} Performance comparison of adding interaction edges for emerging drugs into the training set $\mathcal N_\text{D-train}$.
	Specifically,
	1/3/5 interaction edges in the testing set $\mathcal N_{\text{D-test}}$ are randomly sampled	for each emerging drug in $\mathcal V_{\text{D-test}}$,
	and moved  to the training set $\mathcal N_{\text{D-train}}$.
	\textbf{(c)} Performance comparison of GNN-based methods by varying the depth $L$.
	Specifically,
	$L$ means the number of GNN layers in \texttt{Decagon} and \texttt{DeepLGF},
	the depth of enclosing subgraph in \texttt{SumGNN},
	and the depth of path-based subgraph in \texttt{EmerGNN}.
	\textbf{(d)} Performance comparison of different technique designing in EmerGNN (Supplementary Table~7).}
\label{fig:ablation}
\end{figure}

We compare the performance of top-performing models
according to the frequency of interaction types
to analyze the different models' ability (Fig.~\ref{fig:ablation}a).
\texttt{EmerGNN} outperforms the baselines in all frequencies. 
For the high frequency relations (1\%$\sim$20\%), 
all the methods, except for \texttt{KG-DDI}, have good performance. 
For extremely low frequency relations (81\%$\sim$100\%), 
all the methods work poorly. 
The performance of all methods deteriorates in general for relations
with a lower frequency.
However, the relative performance gain of \texttt{EmerGNN} tends to be larger, especially in the range of 61\%$\sim$80\%. 
These results indicate 
\texttt{EmerGNN}'s strengths in generalization and ability to extract essential information from the biomedical network for predicting rare drugs and interaction types.

The main experiments (Table~\ref{tab:emerging})
study the scenario of emerging drugs without any interaction to existing drugs.
In practice,
we may have a few known interactions between the emerging and existing drugs, often obtained from limited clinical trials.
Hence,
we analyze how different models perform if adding a few interactions for each emerging drug (Fig.~\ref{fig:ablation}b).
We can see that
the performance of shallow models
such as \texttt{CSMDDI} and \texttt{HIN-DNN}
does not change much since the features they use 
are unchanged.
However, methods learning drug embeddings,
such as \texttt{KG-DDI} and \texttt{DeepLGF},
enjoy more substantial improvement when additional knowledge is provided.
In comparison,
\texttt{EmerGNN} has increased performance with more interactions
added and is still the best over all the compared methods.

The value of $L$ determines the maximum number of hops of neighboring entities that the GNN-based models can visit.
We study the impact of changing the length $L$ for these methods
(Fig.~\ref{fig:ablation}c).
The performance of \texttt{Decagon}  and \texttt{DeepLGF} 
gets worse when $L$ gets larger.
Considering that \texttt{Decagon}  and \texttt{DeepLGF}  work on the full biomedical network,
too many irrelevant information will be involved in the representation learning,
leading to worse performance.
\texttt{DeepLGF} runs out-of-memory when $L\geq 3$.
For \texttt{SumGNN} and \texttt{EmerGNN},
$L=1$ performs the worst as the information is hard to be passed from
the emerging drug to the existing drug.
\texttt{SumGNN} can leverage the drug features for prediction,
thus outperforms \texttt{Decagon}.
In comparison,
\texttt{EmerGNN} benefits much 
from the relevant information on the biomedical network
when $L$ increases from 1 to 3.
However,
the performance will decrease when $L>3$.
Intuitively,
the path-based subgraph will contain too much irrelevant information when the length gets longer,
increasing the learning difficulty.
Hence, a moderate number of path length with $L=3$ is optimal for \texttt{EmerGNN},
considering both the effectiveness and computation efficiency.

We conduct experiments to analyze the main techniques in designing
in \texttt{EmerGNN} (Fig.~\ref{fig:ablation}d).
First,
we evaluate the performance of 
using undirected edges without introducing the inverse edges (denoted as Undirected edges w.o. inverse).
It is clear that using undirected edges has negative effect
as the directional information on the biomedical network is lost.
Then,
we design a variant,
that learns a subgraph representation as \texttt{SumGNN}
upon $\mathcal{G}_{u,v}^L$
(denoted as Subgraph representation),
and another variant
that only learns on the uni-directional computing (Method) from direction $u$ to $v$ without considering the direction from $v$ to $u$ (denoted as Uni-directional pair-wise representation).
Comparing subgraph representation with uni-directional pair-wise representation,
we observe that the flow-based GNN architecture is more effective than
the GNN used in \texttt{SumGNN}.
Even though uni-directional pair-wise representation can achieve better performance
compared with all the baselines in S1 setting (Table~\ref{tab:emerging}),
learning bi-directional representations can help to further improve the prediction ability
by balancing the bi-directional communications between drugs.

\section{Discussion}

Predicting drug-drug interactions (DDI) for emerging drugs is a crucial issue in biomedical computational science
as it offers possibilities for treating and alleviating diseases.
Despite recent advances in DDI prediction accuracy
through the use of deep neural networks~\cite{zitnik2018modeling,karim2019drug,tanvir2021predicting,kipf2016semi,gilmer2017neural,ren2022biomedical},
these methods require large amount of known DDI information,
which is often scarce for emerging drugs.
Additionally, some approaches designed for DDI prediction
only leverage shallow features,
limiting their expressiveness in this task.


One limitation of EmerGNN is that the emerging drug to be predicted should be included in the biomedical network.
Building connections between emerging drug and existing drug through 
molecular formula or property may help address this issue.
Although
we demonstrate the effectiveness of EmerGNN for DDI prediction in this paper, 
EmerGNN is a general approach that can be applied to other biomedical applications, 
such as predicting protein-protein interaction, drug-target interaction and disease-gene interaction. 
We anticipate that the paths attended by EmerGNN can enhance the accuracy and interpretability of these predictions.
We hope that our open-sourced EmerGNN can serve as a strong deep learning tool to advance biomedicine and healthcare, 
by enabling practitioners to exploit the rich  knowledge in existing large biomedical networks for low-data scenarios.

\newpage
\section{Methods}
\label{sec11}

To predict interactions between emerging drugs and existing drugs,
it is important to leverage relevant information in
the biomedical network.
Our framework contains four main components:
(i) constructing an augmented network 
by integrating the DDI network with the biomedical network
and adding inverse edges;
(ii) extracting all the paths with length no longer than $L$ from $u$ to $v$ to construct a path-based subgraph $\mathcal{G}_{u,v}^L$;
(iii) encoding pair-wise subgraph representation $\bm h_{u,v}^{(L)}$
by a flow-based GNN with attention mechanism
such that the information can flow from $u$ over the important entities and edges in $\mathcal{G}_{u,v}^L$ to $v$;
(iv) predicting the interaction type based on the bi-directional pair-wise subgraph representations.
The overall framework is shown in Fig.~1.

\paragraph{Augmented network}

Given the DDI network 
$\mathcal N_\text{D}=\{(u,i,v): u,v\in\mathcal V_\text{D}, i\in\mathcal R_\text{I}\}$
and the biomedical network
$\mathcal N_\text{B} = \{(h, r, t): h, t \in \mathcal V_\text{B}, r\in\mathcal R_\text{B}\}$ ($\mathcal N_\text{D}$ is specified as $\mathcal N_\text{D-train}$/$\mathcal N_\text{D-valid}$/$\mathcal N_\text{D-test}$
in the training/validation/testing stages, respectively, 
so does $\mathcal N_\text{B}$),
we integrate the two networks into
\[
\mathcal N' = \mathcal N_\text{D}\cup \mathcal N_\text{B} 
= \big\{(e, r, e'): e, e'=\mathcal V', r\in\mathcal R' \big\},
\]
with $\mathcal V'=\mathcal V_\text{D}\cup \mathcal V_\text{B}$
and $\mathcal R'=\mathcal R_\text{I}\cup \mathcal R_\text{B}$.
The integrated network $\mathcal N'$ connects the existing and emerging drugs by concepts in the biomedical network.
Since the relation types are directed, we follow the common practices in 
knowledge graph learning
\citep{vashishth2019composition,yu2021sumgnn}
to add inverse types.
Specifically,
we add $r_\text{inv}$ for each $r\in\mathcal R'$ and 
create a set of inverse types $\mathcal R'_\text{inv}$,
which subsequently leads to an inverse network
\[\mathcal N'_\text{inv}
= \big\{(e', r_\text{inv}, e): (e,r,e')\in\mathcal N'\big\}.\]
\textcolor{black}{
	Note that the inverse relations will not influence the information in the original
	biomedical network
	since we can transform any inverse edge $(e',r\_\text{inv}, e)$
	back to the original edge $(e,r, e')$.
	Semantically,
	the inverse relations can be regarded as a kind of active voice vs. passive voice in linguistics,
	for instance \textit{includes\_inv} can be regarded as \textit{being included}
	and 
	\textit{causes\_inv} can be regarded as \textit{being caused}.
	By adding the inverse edges,
	the paths can be smoothly organized in single directions.
	For example,
	a path $a\xrightarrow{r1}b\xleftarrow{r2}c$ can be transformed to 
	$a\xrightarrow{r1}b\xrightarrow{r2\_\text{inv}}c$,
	which is more computational friendly.
}

After the above two steps,
we obtain the augmented network
\[
\mathcal N = \mathcal N' \cup \mathcal N'_\text{inv} = \big\{(e,r,e'): e,e'\in\mathcal V, r\in\mathcal R\big\},
\]
with entity set 
$\mathcal V = \mathcal V'=\mathcal V_\text{D}\cup \mathcal V_\text{B}$
and relation set
$\mathcal R = \mathcal R'\cup \mathcal R'_\text{inv}$.

\paragraph{Path-based subgraph formulation}

Inspired by the path-based methods in knowledge graph learning 
\citep{lao2011random,xiong2017deeppath},
we are motivated to extract the paths connecting existing and emerging drugs, 
and predict the interaction type based on the paths.

Given a drug pair $(u,v)$ to be predicted, 
we extract the set $\mathcal P_{u,v}^L$
of all the paths with length up to $L$.
Each path in $\mathcal P_{u,v}^L$ has the form
\[
e_0 \xrightarrow{r_1} e_1 \xrightarrow{r_2} \cdots \xrightarrow{r_L} e_L,
\]
with $e_0 \!=\! u$, 
$e_L\!=\!v$ and $(e_{i-1}$, $r_{i}$, $e_{i})\!\in\!\mathcal N,$ $i \!=\! 1,\dots,L$.
The intermediate entities $e_1,\dots, e_{L-1}\!\in\!\mathcal V$ can be drugs, genes, diseases, side-effects, symptoms, pharmacologic class, etc.,
and $r_1,\dots, r_L\in\mathcal R$ are the interactions or relations between the biomedical entities.
In order preserve the local structures,
we merge the paths in $\mathcal P_{u,v}^L$ to a subgraph $\mathcal{G}_{u,v}^L$
such that the same entities are merged to a single node.
The detailed steps of path extraction and subgraph generation
are provided in Supplementary Section~1.

Different from the subgraph structures used for link prediction
on general graphs
\citep{zhang2018link,teru2020inductive,yu2021sumgnn},
the edges in $\mathcal{G}_{u,v}^L$
are pointed away from $u$ and towards $v$.
Our objective is to learn 
a GNN $g(\cdot)$ with parameters $\bm \theta$ that 
predicts DDI between $u$ and $v$ based on the path-based subgraph $\mathcal{G}_{u,v}^L$, that is
\begin{equation}
	\text{DDI}(u, v) = g\left(\mathcal{G}_{u,v}^L; \bm \theta\right).
	\label{eq:gnn}
\end{equation}
The link prediction problem on the DDI network
is then transformed as a whole graph learning problem.

\paragraph{Flow-based GNN architecture}

Given $\mathcal{G}_{u,v}^L$, we would like to integrate essential information
in it to predict the target interaction type.
Note that the edges in $\mathcal{G}_{u,v}^L$ are from the paths 
$\mathcal P_{u,v}^L$ connecting from $u$ to $v$.
We aim to design a special GNN architecture
that the information can flow from drug $u$  to $v$,
via integrating entities and relations
in $\mathcal{G}_{u,v}^L$.

Denote $\mathcal V_{u,v}^\ell, \ell=0,\dots,L$, as the set of entities 
that can be visited in the $\ell$-th flow step from $u$ (like the four ellipses in $g(\mathcal{G}_{u,v}^L;\bm \theta)$ in Fig.~\ref{fig:pipeline}).
In particular, we have $\mathcal V_{u,v}^0 =\{u\}$ as the starting point
and $\mathcal V_{u,v}^L =\{v\}$ as the ending point.
\textcolor{black}{
	In the $\ell$-th iteration,
	the visible entities in $\mathcal V_{u,v}^\ell$ contains entities
	that are $\ell$-steps away from drug $u$ and 
	are $(L-\ell)$-steps away from drug $v$
	in the augmented network $\mathcal N$.
}
We use the fingerprint features 
\citep{rogers2010extended} of drug $u$ 
as the input representation of
$u$,
namely $\bm h_{u,u}^{(0)}=\bm f_u$.
Then,
we conduct message flow for $L$ steps with the function
\begin{equation}
	\bm h_{u,e}^{(\ell)} 
	\!=\! \delta\!\!\left(\!\bm W^{(\ell)}\!\!\sum\nolimits_{e'\in\mathcal V_{u,v}^{\ell-1}} \!\big(\bm h_{u,e}^{(\ell-1)}\!+\!\phi(\bm h_{u,e'}^{(\ell-1)}\!,\bm h_r^{(\ell)})\big)\!\!\right),
	\label{eq:prop}
\end{equation}
for entities $e\in\mathcal V^{\ell}_{u,v}$,
where 
$\bm W^{(\ell)} \in\mathbb R^{d\times d}$
is a learnable weighting matrix for step $\ell$;
$\bm h_{u,e'}^{(\ell-1)}$ is the pair-wise representation of entity $e'\in\mathcal V_{u,v}^{\ell-1}$;
$r$ is the relation type between $e'$ and $e$;
$\bm h_r^{(\ell)}\in\mathbb R^d$ is the learnable representation with dimension $d$ of $r$ in the $\ell$-th step;
and $\phi(\cdot, \cdot): (\mathbb R^d,\mathbb R^d)\rightarrow \mathbb R^d$ 
is the function combining the two vectors;
and $\delta(\cdot)$ is the  activation function ReLU \citep{nair2010rectified}.

Since the biomedical network is not specially designed for the DDI prediction task,
we need to control the importance of different edges in $\mathcal{G}_{u,v}^L$.
We use a drug-dependent attention weight
for function $\phi(\cdot, \cdot)$.
Specifically,
we design the message function  for each edge $(e',r,e)$ 
during the $l$-th propagation step as
\begin{equation}
	\phi(\bm h_{u,e'}^{(\ell-1)}, \bm h_r^{(\ell)}) 
	= \alpha_{r}^{(\ell)} 
	\cdot 
	\left( 
	\bm h_{u,e'}^{(\ell-1)}\odot \bm h_{r}^{(\ell)}
	\right),
\end{equation}
where $\odot$ is an element-wise dot product of vectors
and $\alpha_{r}^{(\ell)}$ is the attention weight controlling the importance of messages.
We design the  attention weight depending on the edges' relation type
as 
\begin{align*}
	\alpha_{r}^{(\ell)} = \sigma\left((\bm w_{r}^{(\ell)})^\top[\bm f_{u};\bm f_{v}]\right),
\end{align*} 
where the relation weight
$\bm w_{r}^{(\ell)}\in\mathbb R^{2d}$
is multiplied with the fingerprints $[\bm f_{u};\bm f_{v}]\in\mathbb R^{2d}$
of drugs to be predicted and
$\sigma(\cdot)$ is a sigmoid function returning a value in $(0,1)$.

After iterating for $L$ steps,
we can obtain the representation $\bm h_{u,v}^{(L)}$
that encodes the important paths up to length $L$ between  drugs $u$ and $v$.

\paragraph{Objective and training}

In practice,
the interaction types can be symmetric,
for example \#Drug1 and \#Drug2 may have the side effect of headache if used together,
or asymmetric,
for example
\#Drug1 may decrease the analgesic activities of \#Drug2.
Besides,
the emerging drug can appear in either the source (drug $u$)
or target (drug $v$).
We extract the reverse subgraph $\mathcal{G}_{v,u}^L$
and encode it with the same parameters in Equation~\eqref{eq:prop}
to obtain the reverse pair-wise representation $\bm h_{v,u}^{(L)}$.
Then the bi-directional representations are concatenated 
to predict the interaction type with
\begin{equation}
	\bm l(u,v) 
	= \bm{W}_{\text{rel}}
	\left[ 
	\bm h_{u,v}^{(L)};\bm h_{v,u}^{(L)}
	\right].
	\label{eq:logits}
\end{equation}
Here, 
the  transformation matrix $\bm{W}_{\text{rel}}\in\mathbb R^{\mid\mathcal R_\text{I}\mid\times 2d}$ is used
to map the pair-wise representations into prediction logits $\bm l(u,v)$
of the $\mid\!\!\!\mathcal R_\text{I}\!\!\!\mid$ interaction types.
The $i$-th logit $l_i(u,v)$ indicates the plausibility of interaction type $i$
being predicted.
The full algorithm and implementation details of Equation~\eqref{eq:logits}
are provided in Supplementary Section~1.

Since we have two kinds of tasks
that are multi-class (on the DrugBank dataset)
and multi-label (on the TWOSIDES dataset)
interaction predictions,
the training objectives are different.

For DrugBank,
there exists at most one interaction type between two drugs.
Given two drugs $u$ and $v$,
once we obtain the prediction logits $\bm l(u,v)$
of different interaction types,
we use a softmax function to compute the probability of each interaction type,
namely
\[I_i(u,v) = \frac{\exp\big(l_i(u,v)\big)}{\sum_{j\in\mathcal R_\text{I}}\exp\big(l_j(u,v)\big)}.\]
Denote $\bm y(u,v)\in \mathbb R^{\mid\mathcal R_\text{I}\mid}$ as the 
ground-truth indicator of target
interaction type, 
where $y_i(u,v)=1$ if $(u,i,v)\in\mathcal N_\text{D}$,
otherwise zero.
We minimize the following cross-entropy loss to train the model parameters
\begin{equation}
	\mathcal L_{\text{DB}} 
	= - 
	\!\!\!\!\!\!\!\!\!\!\!\!\!
	\sum_{(u,i,v)\in\mathcal N_{\text{D-train}}}
	\!\!\!\!\!\!\!\!\!\!
	y_i(u,v)\log {I}_i(u,v).
	\label{eq:loss-db}
\end{equation}

For TWOSIDES,
there may be multiple interactions between two drugs.
The objective is to predict whether there is an interaction $p$ between two drugs.
Given two drugs $u$, $v$ and the prediction logits 
$\bm l(u,v)$,
we use the sigmoid function
\begin{align*}
	I_i(u,v) = \frac{1}{1 + \exp( - l_i(u,v) )},
\end{align*}
to compute the probability of interaction type $i$.
Different with the multi-class task in DrugBank,
we use the binary cross entropy loss 
\begin{equation}
	\begin{split}
		\mathcal L_{\text{TS}} = - & \!\!\!\!\!\!\!\!\!\sum_{(u,i,v)\in\mathcal N_{\text{D-train}}}\!\!\!\!\!\Big(\!\log \big(I_i(u,v)\big) + \!\!\!\!\!\! 
		\sum_{(u',v')\in\mathcal N_i} \!\!\!\!\!\!\log\big(1-I_i(u',v')\big) \Big),
	\end{split}
	\label{eq:loss-ts}
\end{equation}
where $\mathcal N_i$ is the set of drug pairs that do not have the interaction type $i$.

We use stochastic gradient optimizer Adam \citep{kingma2014adam}
to optimize the model parameters
\begin{align*}
	\bm \theta =
	\left\lbrace
	\bm W_\text{rel},
	\big\{
	\bm W^{(\ell)}, \bm h_r^{(\ell)}, \bm w_{r}^{(\ell)} \}_{\ell=1,\dots, L, r\in\mathcal R}
	\right\rbrace,
\end{align*}
by minimizing loss function in Equation~\eqref{eq:loss-db}
for the DrugBank dataset
or Equation~\eqref{eq:loss-ts} for the TWOSIDES dataset.

\paragraph{Drug-drug interaction network.}
Following \citep{zitnik2018modeling,yu2021sumgnn},
we use two benchmark datasets,
DrugBank \citep{wishart2018drugbank}
and 
TWOSIDES \citep{tatonetti2012data},
as the interaction network $\mathcal N_\text{D}$ (Supplementary Table~1).
When predicting DDIs for emerging drugs,
namely the S1 and S2 settings, we randomly split $\mathcal V_\text{D}$ into three disjoint sets with
$\mathcal V_\text{D} = \mathcal V_\text{D-train}\cup \mathcal V_\text{D-valid}\cup\mathcal V_\text{D-test}$
and 
$\mathcal V_\text{D-train} \cap \mathcal V_\text{D-valid} \cap \mathcal V_\text{D-test}= \emptyset$,
where 
$\mathcal V_\text{D-train}$ is the set of existing drugs used for training,
$\mathcal V_\text{D-valid}$ is the set of emerging drugs for validation,
and 
$\mathcal V_\text{D-test}$ is the set of emerging drugs for testing.
The interaction network for training is defined as $\mathcal N_{\text{D-train}} = \{(u, i, v)\!\in\! \mathcal N_{\text{D}}:u,v\!\in\!\mathcal V_\text{D-train}\}$.

In the S1 setting, we set
\begin{itemize}
	\item $\mathcal N_{\text{D-valid}} = \{(u, i, v)\!\in\! \mathcal N_{\text{D}}:u\!\in\!\mathcal V_\text{D-train}, v\!\in\!\mathcal V_\text{D-valid}\} \cup  \{(u, i, v)\!\in\! \mathcal N_{\text{D}}:u\!\in\!\mathcal V_\text{D-valid}, v\!\in\!\mathcal V_\text{D-train}\}$ as validation samples;
	\item $\mathcal N_{\text{D-test}} = \{(u, i, v)\!\in\!\mathcal N_\text{D}:u\!\in\!(\mathcal V_{\text{D-train}}\cup\mathcal V_\text{D-valid}), v\!\in\! \mathcal V_{\text{D-test}}\}
	\cup 
	\{(u, i, v)\!\in\!\mathcal N_\text{D}:u\!\in\!\mathcal D_{\text{D-test}}, v\!\in\! (\mathcal V_{\text{D-train}}\cup \mathcal V_\text{D-valid})\}$
	as testing samples.
\end{itemize}

In the S2 setting, we set
\begin{itemize}
	\item $\mathcal N_{\text{D-valid}} = \{(u, i, v)\!\in\! \mathcal N_{\text{D}}:u,v\!\in\!\mathcal V_\text{D-valid}\}$ as validation samples; and 
	\item $\mathcal N_{\text{D-test}} = \{(u, i, v)\!\in\! \mathcal N_{\text{D}}:u,v\!\in\!\mathcal V_\text{D-test}\}$ as testing samples.
\end{itemize}

We follow \cite{yu2021sumgnn}
to randomly sample one negative sample for each $(u,i,v)\in \mathcal N_\text{D-valid}\cup\mathcal N_\text{D-test}$ to form the negative set $\mathcal N_i$
for TWOSIDES dataset
in the evaluation phase.
Specifically,
if $u$ is an emerging drug, we randomly sample an existing drug
$v'\in\mathcal V_\text{D-train}$
and make sure that the new interaction does not exist,
namely $(u,i,v')\notin\mathcal N_\text{D}$;
if $v$ is an emerging drug, we randomly sample an existing drug
$u'\in\mathcal V_\text{D-train}$
and make sure that the new interaction does not exist,
namely $(u',i,v)\notin\mathcal N_\text{D}$.


\paragraph{Biomedical network.}
In this work,
same as the DDI network,
we use different variants of the biomedical network $\mathcal N_\text{B}$
for training, validation and testing.
The well-built biomedical network HetioNet \citep{himmelstein2017systematic}
is used here.
Denote $\mathcal V_\text{B}, \mathcal R_\text{B}, \mathcal N_\text{B}$ as
the set of entities, relations and edges, respectively, in the full biomedical network.
When predicting interactions between existing drugs in the S0 setting,
all the edges in $\mathcal N_\text{B}$ are used for training, validation and testing.
When predicting interactions between emerging drugs and existing drugs (S1 and S2 setting),
we use different parts of the biomedical networks.

In order to guarantee that the emerging drugs are connected 
with some existing drugs through the biomedical entities,
we constrain the split of drugs to satisfy the conditions
$\mathcal V_\text{D-valid}\subset \mathcal V_\text{B}$ and 
$\mathcal V_\text{D-test}\subset \mathcal V_\text{B}$.
Meanwhile,
we also guarantee that the emerging drugs will not be seen
in the biomedical network during training.
To achieve this goal,
the edges for training are in the set
$\mathcal N_\text{B-train} = \{(h,r,t)\in\mathcal N_\text{B}: h, t\notin(\mathcal V_\text{D-valid}\cup\mathcal V_\text{D-test})\}$;
the edges for validation are in the set
$\mathcal N_\text{B-valid} = \{(h,r,t)\in\mathcal N_\text{B}: h, t\notin\mathcal V_\text{D-test} \}$;
and the testing network is the original network,
namely
$\mathcal N_\text{B-test} = \mathcal N_\text{B}$.

%

{\color{black}
	In addition,
	we plot the size distribution (measured by the number of edges in $\mathcal G_{u,v}^L$)
	as histograms (Supplementary Figure~1).
	We observe that both datasets follow long-tail distributions.
	Many subgraphs have tens of thousands of edges
	on DrugBank,
	while hundreds of thousands of  edges on TWOSIDES since the DDI network is denser.
	Comparing with the augmented networks,
	whose sizes are 3,657,114 for DrugBank 
	and 3,567,059 for TWOSIDES,
	the sizes of subgraphs are quite small.}

\paragraph{Evaluation metrics.}
As pointed by \cite{yu2021sumgnn},
there is at most one interaction between a pair of drugs in the DrugBank dataset \citep{wishart2018drugbank}.
Hence, we evaluate the performance in a multi-class setting,
which estimates whether the model can correctly predict the interaction type for a pair of drugs.
We consider the following metrics:
\begin{itemize}
	\item $\text{F1(macro)}=\frac{1}{\|\mathcal I_D\|}\sum_{i\in\mathcal I_D}\frac{2P_i\cdot R_i}{P_i+R_i}$,
	where $P_i$ and $R_i$ are the precision and recall for the interaction type $i$, respectively.
	The macro F1 aggregates the fractions over different interaction types.
	\item Accuracy: the percentage of correctly predicted interaction type compared with the ground-truth interaction type.
	\item Cohen's Kappa \citep{cohen1960coefficient}: $\kappa = \frac{A_p-A_e}{1-A_e}$, where $A_p$ is the observed agreement (accuracy) and $A_e$ is the probability of randomly seeing each class.
\end{itemize}

In the TWOSIDES dataset \citep{tatonetti2012data},
there may be multiple interactions between a pair of drugs, such as anaemia, nausea and pain.
Hence, we model and evaluate the performance in a multi-label setting,
where each type of side effect is modeled as a binary classification problem.
Following \citep{zitnik2018modeling,tatonetti2012data},
we sample one negative drug pair for each $(u,i,v)\in\mathcal N_{\text{D-test}}$ 
and evaluate the binary classification performance with the following metrics:
\begin{itemize}
	\item ROC-AUC: the area under curve of receiver operating characteristics,
	measured by $\sum_{k=1}^n\text{TP}_k\Delta \text{FP}_k$, 
	where $(\text{TP}_k, \text{FP}_k)$ 
	is the true-positive and false-positive of the $k$-th operating point.
	\item PR-AUC: the area under curve of precision-recall,
	measured by $\sum_{k=1}^n{P}_k\Delta {R}_k$, 
	where $({P}_k,{R}_k)$ is the precision and recall of the $k$-th operating point.
	\item Accuracy: the average precision of drug pairs for each side effect.
\end{itemize}

\section*{Data availability}
Source data for Figures 2-5 is available with this manuscript.
The resplit dataset \cite{zhang_2023_10016715} of DrugBank, TWOSIDES and HetioNet 
for S1 and S2 settings is public available
at \url{https://doi.org/10.5281/zenodo.10016715}.


\section*{Code availability}
The code for EmerGNN \cite{Zhang_EmerGNN}  is available
at \url{https://github.com/LARS-research/EmerGNN}.

\section*{Acknowledgments}

This project was supported by the National Natural Science Foundation of China (No. 92270106)
and CCF-Tencent Open Research Fund.

\section*{Author Contributions Statement}

Y. Zhang contributes to 
idea development,
algorithm implementation,
experimental design,
result analysis,
and paper writing.
Q. Yao contributes to
idea development,
experimental design,
result analysis,
and paper writing.
L. Yue 
contributes to
algorithm implementation
and 
result analysis.
Y. Zheng
contributes to
result analysis
and paper writing.
All 
authors read, edited, and approved the paper.

\section*{Competing Interests Statement}

The authors declare no competing interests.

\newpage

\section*{Tables}

\begin{table}[H]
	\centering
	\caption{Performance of EmerGNN compared other DDI prediction methods.
		Four types of DDI prediction methods are compared:
		(i) methods that use drug features of target drug pairs (\textit{DF})~\cite{rogers2010extended,vilar2014similarity,liu2022predict,yu2022stnn};
		(ii) methods that use graph features in the biomedical network (\textit{GF})~\cite{tanvir2021predicting};
		(iii) methods that learn drug embeddings (\textit{Emb})~\cite{yao2022effective,karim2019drug};
		and
		(iv) methods that model with GNNs (\textit{GNN})~\cite{vashishth2019composition,zitnik2018modeling,lin2020kgnn,yu2021sumgnn,ren2022biomedical}.
	}
	\label{tab:emerging}
	\vspace{3px}
	
	\setlength\tabcolsep{3pt}
	\renewcommand\arraystretch{0.9}
	\leftline{\small \textbf{S1 Setting}: DDI prediction between emerging drug and existing drug
		\tablefootnote{
			All of the methods are run for five times on the five-fold datasets
			with mean value and standard deviation reported on the testing data.
			The evaluation metrics are presented in percentage (\%)
			with the larger value indicating better performance.
			The boldface numbers indicate the best values,
			while the underlined numbers indicate the second best.
			p-values are computed under two-sided t-testing of 
			EmerGNN over the second best baselines.
			Methods leveraging a biomedical network are indicated by star *.}.
	}
	\resizebox{\textwidth}{!}{
		\begin{tabular}{ll|ccc|ccc}
			\toprule
			\multicolumn{2}{c|}{Datasets}           &                     \multicolumn{3}{c|}{\textbf{DrugBank}}                     &                     \multicolumn{3}{c}{\textbf{TWOSIDES}}                      \\ \midrule
			Type & Methods                                   &         F1-Score         &         Accuracy         &         Kappa         &          PR-AUC        &         ROC-AUC            &         Accuracy         \\ \midrule
			DF   & MLP \citep{rogers2010extended}      &   21.1$\pm$0.8   &   46.6$\pm$2.1     &    33.4$\pm$2.5     &   81.5$\pm$1.5  &   81.2$\pm$1.9      &  76.0$\pm$2.1    \\
			& Similarity \citep{vilar2014similarity}       &  43.0$\pm$5.0 & 51.3$\pm$3.5 & 44.8$\pm$3.8  &   56.2$\pm$0.5  &      55.7$\pm$0.6    &    53.9$\pm$0.4      \\  	
			& CSMDDI \citep{liu2022predict} & \underline{45.5}$\pm$1.8  & \underline{62.6}$\pm$2.8 & \underline{55.0}$\pm$3.2  &   73.2$\pm$2.6 &  74.2$\pm$2.9  &  69.9$\pm$2.2  \\
			&  STNN-DDI \citep{yu2022stnn} & 39.7$\pm$1.8  &  56.7$\pm$2.6	&  46.5$\pm$3.4  & 68.9$\pm$2.0  & 68.3$\pm$2.6  &  65.3$\pm$1.8 \\
			\midrule
			GF   & HIN-DDI* \citep{tanvir2021predicting}        &  37.3$\pm$2.9    &   58.9$\pm$1.4    &   47.6$\pm$1.8    & 	\underline{81.9}$\pm$0.6  &	\underline{83.8}$\pm$0.9	& \underline{79.3}$\pm$1.1	\\	\midrule
			Emb  & MSTE \citep{yao2022effective}     &   7.0$\pm$0.7   &     51.4$\pm$1.8     &    37.4$\pm$2.2       &     64.1$\pm$1.1     &     62.3$\pm$1.1   &  58.7$\pm$0.7     \\
			& KG-DDI* \citep{karim2019drug}     & 26.1$\pm$0.9  &  46.7$\pm$1.9 &    35.2$\pm$2.5 &     79.1$\pm$0.9   &       77.7$\pm$1.0     &   60.2$\pm$2.2       \\	 \midrule
			GNN  & CompGCN*  \citep{vashishth2019composition}    &    26.8$\pm$2.2    &     48.7$\pm$3.0     &     37.6$\pm$2.8      &     80.3$\pm$3.2    &    79.4$\pm$4.0    &    71.4$\pm$3.1      \\
			& Decagon* \citep{zitnik2018modeling}          &    24.3$\pm$4.5   &    47.4$\pm$4.9        &     35.8$\pm$5.9      &    79.0$\pm$2.0  &   78.5$\pm$2.3   &     69.7$\pm$2.4     \\
			& KGNN* \citep{lin2020kgnn}                   & 23.1$\pm$3.4   &   51.4$\pm$1.9    &    40.3$\pm$2.7       &    78.5$\pm$0.5   &    79.8$\pm$0.6    &   72.3$\pm$0.7       \\
			& SumGNN* \citep{yu2021sumgnn}               &   35.0$\pm$4.3    &   48.8$\pm$8.2         &      41.1$\pm$4.7     &     80.3$\pm$1.1    &   81.4$\pm$1.0         &       73.0$\pm$1.4   \\
			& DeepLGF* \citep{ren2022biomedical}  &  39.7$\pm$2.3  &    60.7$\pm$2.4       &       51.0$\pm$2.6     &   81.4$\pm$2.1      &      82.2$\pm$2.6      &   72.8$\pm$2.8     \\
			& \textbf{EmerGNN}*               &    \textbf{62.0}$\pm$2.0        &   \textbf{68.6}$\pm$3.7    &   \textbf{62.4}$\pm$4.3       &  \textbf{90.6}$\pm$0.7  &   \textbf{91.5}$\pm$1.0    &    \textbf{84.6}$\pm$0.7    \\	\midrule
			\multicolumn{2}{l|}{\textbf{p-value}}      &    8.9E-7    &    0.02        &     0.02     &      1.6E-6   &    6.0E-8  &   3.5E-5       \\ \bottomrule
		\end{tabular}
	}
	
	\vspace{8px}
	\leftline{\small \textbf{S2 Setting}: DDI prediction between two emerging drugs.}
	\resizebox{\textwidth}{!}{
		\begin{tabular}{ll|ccc|ccc}    
			\toprule
			\multicolumn{2}{c|}{Datasets}            &                     \multicolumn{3}{c|}{\textbf{DrugBank}}                     &                     \multicolumn{3}{c}{\textbf{TWOSIDES}}                      \\ \midrule
			Type & 	Methods                                   &         F1-Score         &         Accuracy         &         Kappa         &          PR-AUC        &         ROC-AUC        &         Accuracy         \\ \midrule
			DF  &	CSMDDI~\citep{liu2022predict}  	 &  \underline{19.8}$\pm$3.1    &   \underline{37.3}$\pm$4.8      &   \underline{22.0}$\pm$4.9     &      55.8$\pm$4.9       &    57.0$\pm$6.1   &      55.1$\pm$5.2    \\ \midrule
			GF &	HIN-DDI*~\citep{tanvir2021predicting}         &   8.8$\pm$1.0   &    27.6$\pm$2.4     &   13.8$\pm$2.4    &    \underline{64.8}$\pm$2.3     &  \underline{58.5}$\pm$1.6     &   \underline{59.8}$\pm$1.4       \\ \midrule
			Emb &	KG-DDI*~\citep{karim2019drug}     &   1.1$\pm0.1$    &    32.2$\pm$3.6     &    0.0$\pm$0.0   &   53.9$\pm$3.9   &   47.0$\pm$5.5    &    50.0$\pm$0.0      \\	 \midrule
			GNN&	DeepLGF*~\citep{ren2022biomedical}    &    4.8$\pm$1.9   &    31.9$\pm$3.7     &   8.2$\pm$2.3     &     59.4$\pm$8.7   &   54.7$\pm$5.9      &    54.0$\pm$6.2      \\ 
			&	\textbf{EmerGNN}*               &   \textbf{25.0}$\pm$2.8   &   \textbf{46.3}$\pm$3.6    &     \textbf{31.9}$\pm$3.8       &    \textbf{81.4}$\pm$7.4   &    \textbf{79.6}$\pm$7.9   &  \textbf{73.0}$\pm$8.2     \\  \midrule
			\multicolumn{2}{l|}{\textbf{p-value}}      &   0.02   &  0.01  &    0.01   &    1.4E-3     &      3.9E-4      &     7.8E-3     \\ \bottomrule
		\end{tabular}
	}
\end{table}

\section*{Figures}

\newpage

\bibliographystyle{unsrtnat}
\bibliography{document}

\begin{thebibliography}{40}
\providecommand{\natexlab}[1]{#1}
\providecommand{\url}[1]{\texttt{#1}}
\expandafter\ifx\csname urlstyle\endcsname\relax
  \providecommand{\doi}[1]{doi: #1}\else
  \providecommand{\doi}{doi: \begingroup \urlstyle{rm}\Url}\fi

\bibitem[Su et~al.(2022)Su, Wang, Zhao, Wang, and Cui]{su2022trends}
Xian Su, Haixue Wang, Nan Zhao, Tao Wang, and Yimin Cui.
\newblock Trends in innovative drug development in {China}.
\newblock \emph{Nature Reviews. Drug Discovery}, 2022.

\bibitem[Ledford(2022)]{ledford2022hundreds}
Heidi Ledford.
\newblock Hundreds of {COVID} trials could provide a deluge of new drugs.
\newblock \emph{Nature}, pages 25--27, 2022.

\bibitem[Percha and Altman(2013)]{percha2013informatics}
Bethany Percha and Russ~B Altman.
\newblock Informatics confronts drug-drug interactions.
\newblock \emph{Trends in Pharmacological Sciences}, 34\penalty0 (3):\penalty0
  178--184, 2013.

\bibitem[Vilar et~al.(2014)Vilar, Uriarte, Santana, Lorberbaum, Hripcsak,
  Friedman, and Tatonetti]{vilar2014similarity}
Santiago Vilar, Eugenio Uriarte, Lourdes Santana, Tal Lorberbaum, George
  Hripcsak, Carol Friedman, and Nicholas~P Tatonetti.
\newblock Similarity-based modeling in large-scale prediction of drug-drug
  interactions.
\newblock \emph{Nature Protocols}, 9\penalty0 (9):\penalty0 2147--2163, 2014.

\bibitem[Tanvir et~al.(2021)Tanvir, Islam, and Akbas]{tanvir2021predicting}
Farhan Tanvir, Muhammad Ifte~Khairul Islam, and Esra Akbas.
\newblock Predicting drug-drug interactions using meta-path based similarities.
\newblock In \emph{IEEE Conference on Computational Intelligence in
  Bioinformatics and Computational Biology}, pages 1--8. IEEE, 2021.

\bibitem[Yu et~al.(2021)Yu, Huang, Zhang, Glass, Sun, and Xiao]{yu2021sumgnn}
Yue Yu, Kexin Huang, Chao Zhang, Lucas~M Glass, Jimeng Sun, and Cao Xiao.
\newblock {SumGNN}: multi-typed drug interaction prediction via efficient
  knowledge graph summarization.
\newblock \emph{Bioinformatics}, 37\penalty0 (18):\penalty0 2988--2995, 2021.

\bibitem[Letinier et~al.(2019)Letinier, Cossin, Mansiaux, Arnaud, Salvo, Bezin,
  Thiessard, and Pariente]{Letinier2019}
Louis Letinier, Sebastien Cossin, Yohann Mansiaux, Mickael Arnaud, Francesco
  Salvo, Julien Bezin, Frantz Thiessard, and Antoine Pariente.
\newblock Risk of drug-drug interactions in out-hospital drug dispensings in
  {France}: Results from the drug-drug interaction prevalence study.
\newblock \emph{Frontiers in Pharmacology}, 10:\penalty0 265, 2019.

\bibitem[Jiang et~al.(2022)Jiang, Lin, Ren, Fang, Liu, Tan, Lv, and
  Zhang]{Jiang2022}
Huaqiao Jiang, Yanhua Lin, Weifang Ren, Zhonghong Fang, Yujuan Liu, Xiaofang
  Tan, Xiaoqun Lv, and Ning Zhang.
\newblock Adverse drug reactions and correlations with drug-drug interactions:
  A retrospective study of reports from 2011 to 2020.
\newblock \emph{Frontiers in Pharmacology}, 13, 2022.

\bibitem[Rogers and Hahn(2010)]{rogers2010extended}
David Rogers and Mathew Hahn.
\newblock Extended-connectivity fingerprints.
\newblock \emph{Journal of Chemical Information and Modeling}, 50\penalty0
  (5):\penalty0 742--754, 2010.

\bibitem[Dewulf et~al.(2021)Dewulf, Stock, and De~Baets]{dewulf2021cold}
Pieter Dewulf, Michiel Stock, and Bernard De~Baets.
\newblock Cold-start problems in data-driven prediction of drug-drug
  interaction effects.
\newblock \emph{Pharmaceuticals}, 14\penalty0 (5):\penalty0 429, 2021.

\bibitem[Liu et~al.(2022)Liu, Wang, Yu, Shi, and Dong]{liu2022predict}
Zun Liu, Xing-Nan Wang, Hui Yu, Jian-Yu Shi, and Wen-Min Dong.
\newblock Predict multi-type drug-drug interactions in cold start scenario.
\newblock \emph{BMC Bioinformatics}, 23\penalty0 (1):\penalty0 75, 2022.

\bibitem[Yao et~al.(2022)Yao, Sun, Jian, Wu, and Wang]{yao2022effective}
Junfeng Yao, Wen Sun, Zhongquan Jian, Qingqiang Wu, and Xiaoli Wang.
\newblock Effective knowledge graph embeddings based on multidirectional
  semantics relations for polypharmacy side effects prediction.
\newblock \emph{Bioinformatics}, 38\penalty0 (8):\penalty0 2315--2322, 2022.

\bibitem[Zitnik et~al.(2018)Zitnik, Agrawal, and Leskovec]{zitnik2018modeling}
Marinka Zitnik, Monica Agrawal, and Jure Leskovec.
\newblock Modeling polypharmacy side effects with graph convolutional networks.
\newblock \emph{Bioinformatics}, 34\penalty0 (13):\penalty0 i457--i466, 2018.

\bibitem[Karim et~al.(2019)Karim, Cochez, Jares, Uddin, Beyan, and
  Decker]{karim2019drug}
Md~Rezaul Karim, Michael Cochez, Joao~Bosco Jares, Mamtaz Uddin, Oya Beyan, and
  Stefan Decker.
\newblock Drug-drug interaction prediction based on knowledge graph embeddings
  and convolutional-{LSTM} network.
\newblock In \emph{Proceedings of the 10th ACM International Conference on
  Bioinformatics, Cmputational Biology and Health Informatics}, pages 113--123,
  2019.

\bibitem[Huang et~al.(2020)Huang, Xiao, Glass, Zitnik, and
  Sun]{huang2020skipgnn}
Kexin Huang, Cao Xiao, Lucas~M Glass, Marinka Zitnik, and Jimeng Sun.
\newblock {SkipGNN}: predicting molecular interactions with skip-graph
  networks.
\newblock \emph{Scientific Reports}, 10\penalty0 (1):\penalty0 1--16, 2020.

\bibitem[Lin et~al.(2020)Lin, Quan, Wang, Ma, and Zeng]{lin2020kgnn}
Xuan Lin, Zhe Quan, Zhi-Jie Wang, Tengfei Ma, and Xiangxiang Zeng.
\newblock {KGNN}: Knowledge graph neural network for drug-drug interaction
  prediction.
\newblock In \emph{Proceedings of the Twenty-Ninth International Conference on
  International Joint Conferences on Artificial Intelligence}, volume 380,
  pages 2739--2745, 2020.

\bibitem[Ren et~al.(2022)Ren, You, Yu, Li, Guan, Guo, and
  Pan]{ren2022biomedical}
Zhong-Hao Ren, Zhu-Hong You, Chang-Qing Yu, Li-Ping Li, Yong-Jian Guan,
  Lu-Xiang Guo, and Jie Pan.
\newblock A biomedical knowledge graph-based method for drug-drug interactions
  prediction through combining local and global features with deep neural
  networks.
\newblock \emph{Briefings in Bioinformatics}, 23\penalty0 (5):\penalty0
  bbac363, 2022.

\bibitem[Himmelstein et~al.(2017)Himmelstein, Lizee, Hessler, Brueggeman, Chen,
  Hadley, Green, Khankhanian, and Baranzini]{himmelstein2017systematic}
Daniel~Scott Himmelstein, Antoine Lizee, Christine Hessler, Leo Brueggeman,
  Sabrina~L Chen, Dexter Hadley, Ari Green, Pouya Khankhanian, and Sergio~E
  Baranzini.
\newblock Systematic integration of biomedical knowledge prioritizes drugs for
  repurposing.
\newblock \emph{Elife}, 6:\penalty0 e26726, 2017.

\bibitem[Kipf and Welling(2016)]{kipf2016semi}
Thomas~N Kipf and Max Welling.
\newblock Semi-supervised classification with graph convolutional networks.
\newblock In \emph{International Conference on Learning Representations}, 2016.

\bibitem[Gilmer et~al.(2017)Gilmer, Schoenholz, Riley, Vinyals, and
  Dahl]{gilmer2017neural}
Justin Gilmer, Samuel~S Schoenholz, Patrick~F Riley, Oriol Vinyals, and
  George~E Dahl.
\newblock Neural message passing for quantum chemistry.
\newblock In \emph{International Conference on Machine Learning}, pages
  1263--1272. PMLR, 2017.

\bibitem[Yu et~al.(2022)Yu, Zhao, and Shi]{yu2022stnn}
Hui Yu, ShiYu Zhao, and JianYu Shi.
\newblock {STNN}-{DDI}: a substructure-aware tensor neural network to predict
  drug-drug interactions.
\newblock \emph{Briefings in Bioinformatics}, 23\penalty0 (4):\penalty0
  bbac209, 2022.

\bibitem[Wishart et~al.(2018)Wishart, Feunang, Guo, Lo, Marcu, Grant, Sajed,
  Johnson, Li, Sayeeda, et~al.]{wishart2018drugbank}
David~S Wishart, Yannick~D Feunang, An~C Guo, Elvis~J Lo, Ana Marcu, Jason~R
  Grant, Tanvir Sajed, Daniel Johnson, Carin Li, Zinat Sayeeda, et~al.
\newblock {DrugBank} 5.0: a major update to the {DrugBank} database for 2018.
\newblock \emph{Nucleic Acids Research}, 46\penalty0 (D1):\penalty0
  D1074--D1082, 2018.

\bibitem[Tatonetti et~al.(2012)Tatonetti, Ye, Daneshjou, and
  Altman]{tatonetti2012data}
Nicholas~P Tatonetti, Patrick~P Ye, Roxana Daneshjou, and Russ~B Altman.
\newblock Data-driven prediction of drug effects and interactions.
\newblock \emph{Science translational medicine}, 4\penalty0 (125):\penalty0
  125ra31--125ra31, 2012.

\bibitem[Cohen(1960)]{cohen1960coefficient}
Jacob Cohen.
\newblock A coefficient of agreement for nominal scales.
\newblock \emph{Educational and Psychological Measurement}, 20\penalty0
  (1):\penalty0 37--46, 1960.

\bibitem[Brown et~al.(2021)Brown, Wobst, Kapoor, Kenna, and
  Southall]{brown2021clinical}
Dean~G Brown, Heike~J Wobst, Abhijeet Kapoor, Leslie~A Kenna, and Noel
  Southall.
\newblock Clinical development times for innovative drugs.
\newblock \emph{Nature reviews. Drug discovery}, 21\penalty0 (11):\penalty0
  793--794, 2021.

\bibitem[Liu and Wittbrodt(2002)]{liu2002low}
Maywin Liu and Eric Wittbrodt.
\newblock Low-dose oral naloxone reverses opioid-induced constipation and
  analgesia.
\newblock \emph{Journal of Pain and Symptom Management}, 23\penalty0
  (1):\penalty0 48--53, 2002.

\bibitem[Estabrook(2003)]{estabrook2003passion}
Ronald~W Estabrook.
\newblock A passion for {P450s} (remembrances of the early history of research
  on cytochrome {P450}).
\newblock \emph{Drug Metabolism and Disposition}, 31\penalty0 (12):\penalty0
  1461--1473, 2003.

\bibitem[Vashishth et~al.(2019)Vashishth, Sanyal, Nitin, and
  Talukdar]{vashishth2019composition}
Shikhar Vashishth, Soumya Sanyal, Vikram Nitin, and Partha Talukdar.
\newblock Composition-based multi-relational graph convolutional networks.
\newblock In \emph{International Conference on Learning Representations}, 2019.

\bibitem[Lao et~al.(2011)Lao, Mitchell, and Cohen]{lao2011random}
Ni~Lao, Tom Mitchell, and William Cohen.
\newblock Random walk inference and learning in a large scale knowledge base.
\newblock In \emph{Proceedings of the 2011 Conference on Empirical Methods in
  Natural Language Processing}, pages 529--539, 2011.

\bibitem[Xiong et~al.(2017)Xiong, Hoang, and Wang]{xiong2017deeppath}
Wenhan Xiong, Thien Hoang, and William~Yang Wang.
\newblock {DeepPath}: A reinforcement learning method for knowledge graph
  reasoning.
\newblock In \emph{Proceedings of the 2017 Conference on Empirical Methods in
  Natural Language Processing}, pages 564--573, 2017.

\bibitem[Zhang and Chen(2018)]{zhang2018link}
Muhan Zhang and Yixin Chen.
\newblock Link prediction based on graph neural networks.
\newblock In \emph{Proceedings of the 32nd International Conference on Neural
  Information Processing Systems}, pages 5171--5181, 2018.

\bibitem[Teru et~al.(2020)Teru, Denis, and Hamilton]{teru2020inductive}
Komal Teru, Etienne Denis, and Will Hamilton.
\newblock Inductive relation prediction by subgraph reasoning.
\newblock In \emph{International Conference on Machine Learning}, pages
  9448--9457. PMLR, 2020.

\bibitem[Nair and Hinton(2010)]{nair2010rectified}
Vinod Nair and Geoffrey~E Hinton.
\newblock Rectified linear units improve restricted {Boltzmann} machines.
\newblock In \emph{Proceedings of the 27th International Conference on Machine
  Learning}, pages 807--814, 2010.

\bibitem[Kingma and Ba(2014)]{kingma2014adam}
D.~P Kingma and J.~Ba.
\newblock Adam: A method for stochastic optimization.
\newblock Technical report, arXiv:1412.6980, 2014.

\bibitem[Zhang et~al.(2023{\natexlab{a}})Zhang, Yue, and
  Yao]{zhang_2023_10016715}
Yongqi Zhang, Ling Yue, and Quanming Yao.
\newblock {EmerGNN\_DDI\_data}, October 2023{\natexlab{a}}.

\bibitem[Zhang et~al.(2023{\natexlab{b}})Zhang, Yue, and Yao]{Zhang_EmerGNN}
Yongqi Zhang, Ling Yue, and Quanming Yao.
\newblock {EmerGNN}, 10 2023{\natexlab{b}}.
\newblock URL \url{https://github.com/LARS-research/EmerGNN}.

\bibitem[Van~der Maaten and Hinton(2008)]{van2008visualizing}
Laurens Van~der Maaten and Geoffrey Hinton.
\newblock Visualizing data using t-{SNE}.
\newblock \emph{Journal of Machine Learning Research}, 9\penalty0 (11), 2008.

\bibitem[Bergstra et~al.(2015)Bergstra, Komer, Eliasmith, Yamins, and
  Cox]{bergstra2015hyperopt}
James Bergstra, Brent Komer, Chris Eliasmith, Dan Yamins, and David~D Cox.
\newblock Hyperopt: A {Python} library for model selection and hyperparameter
  optimization.
\newblock \emph{Computational Science \& Discovery}, 8\penalty0 (1):\penalty0
  014008, 2015.

\bibitem[Chaudhry et~al.(2022)Chaudhry, Miao, and
  Rehman]{chaudhry2022physiology}
Raheel Chaudhry, Julia~H Miao, and Afzal Rehman.
\newblock Physiology, cardiovascular.
\newblock In \emph{StatPearls [Internet]}. StatPearls Publishing, 2022.

\bibitem[Xu et~al.(2018)Xu, Hu, Leskovec, and Jegelka]{xu2018powerful}
Keyulu Xu, Weihua Hu, Jure Leskovec, and Stefanie Jegelka.
\newblock How powerful are graph neural networks?
\newblock In \emph{International Conference on Learning Representations}, 2018.

\end{thebibliography}

\setcounter{table}{0}
\setcounter{figure}{0}
\setcounter{section}{0}

\makeatletter
\renewcommand{\tablename}{Supplementary Table~\@gobble}
\renewcommand{\figurename}{Supplementary Figure~\@gobble}
\makeatother

\renewcommand\thesection{Supplementary Section~\arabic{section}}

\clearpage
\section{Algorithms}

\paragraph{Algorithm for EmerGNN.}
In this part, we show the full algorithm and some implementation details of EmerGNN.
Given the augmented network $\mathcal N$ and the drug pairs $(u,v)$,
it will be time consuming to explicitly extract all the paths connecting $u$ and $v$ with length $\leq L$.
In practice, we implicitly encode the pair-wise representations with Algorithm~\ref{alg:emergnn}.

\begin{algorithm}[H]
	\caption{EmerGNN: pair-wise representation learning with flow-based GNN.}
	\label{alg:emergnn}
	\begin{algorithmic}[1]
		\REQUIRE {$(u, v), L, \delta, \sigma, \{\bm W^{(\ell)}, \bm w^{(\ell)}\}_{\ell=1\dots L}\}$.\\} \COMMENT{$(u, v)$: drug pair; $L$: the  depth of path-based subgraph; $\delta$: activation function; $\sigma$: sigmoid function; $\{\bm W^{(\ell)}, \bm w^{(\ell)}\}_{\ell=1\dots L}\}$: learnable parameters.}
		\STATE initialize the $u\rightarrow v$ pair-wise representation as $\bm h_{u,e}^0 = \bm f_u$ if $e=u$, otherwise $\bm h_{u,e}^0 = \bm 0$;
		\STATE initialize the $v\rightarrow u$ pair-wise representation as $\bm h_{v,e}^0 = \bm f_v$ if $e=v$, otherwise $\bm h_{v,e}^0 = \bm 0$;
		\FOR{$\ell\leftarrow 1$ to $L$}
		\FOR[This loop can work with matrix operations in parallel.]{$e\in\mathcal V_\text{D}$}
		\STATE message for $u\rightarrow v$: \\$\bm h_{u,e}^{(\ell)} 
		= \delta\left(\bm W^{(\ell)}\sum_{(e',r,e)\in\mathcal N_\text{D}} \!
		\sigma\left((\bm w_r^{(\ell)})^\top [\bm f_u; \bm f_v]\right)\cdot \left(\bm h_{u,e'}^{(\ell-1)}\odot\bm h_{r}^{(\ell)} \right)  \right)$;
		\STATE message for $v\rightarrow u$: \\$\bm h_{v,e}^{(\ell)} 
		= \delta\left(\bm W^{(\ell)}\sum_{(e',r,e)\in\mathcal N_\text{D}} \!
		\sigma\left((\bm w_r^{(\ell)})^\top [\bm f_u; \bm f_v]\right)\cdot \left(\bm h_{v,e'}^{(\ell-1)}\odot\bm h_{r}^{(\ell)} \right)  \right)$;
		\ENDFOR
		\ENDFOR
		\STATE \textbf{Return} $\bm W_\text{rel}[\bm h_{u,v}^{(L)}; \bm h_{v,u}^{(L)}]$.
	\end{algorithmic}
\end{algorithm}

Take the direction $u\rightarrow v$ as an example.
We initialize the representation
$\bm h_{u,e}^0 = \bm f_u$ if $e=u$, otherwise $\bm h_{u,e}^0 = \bm 0$
and the messages are computed based on a dot product operator 
$\bm h_{u,e'}^{(\ell-1)}\odot\bm h_{r}^{(\ell)} $,
then the representations of all entities with length longer than $\ell$ away from $u$
will be $\bm 0$ in the $\ell$-th step.
In the end,
only the entities with length $\leq L$ will have valid representations.
In addition,
since we return $\bm h_{u,v}^{(L)}$ for specific entity $v$,
only the entities with length less than $L-\ell$ away from $v$ can contribute to $\bm h_{u,v}^{(L)}$
in the $\ell$-th step.
In this way,
we implicitly encode relevant entities and relations
in the biomedical network from $u$ to $v$.
We provide a graphical illustration (Supplementary Figure~\ref{fig:illustration}) of the implicit encoding procedure as follows.
\begin{itemize}[leftmargin=*]
	\item When $\ell=0$, only $\bm h_{u,u}^0$ is initialized with the non-zero features $\bm f_{u}$ (in black) and  other entities are initialized as $\mathbf 0$ (in gray).
	\item During the $\ell$-th iteration, the representations are flowed from $u$
	to the $\ell$-th hop neighbors of $u$ in the $\ell$-th step (like the formulas in black, representing the representing a node in corresponding layer).
	\item At the last step $\ell=L$, $\bm h_{u,v}^L$ is used as the subgraph representation. We use boxes to indicate the representations participated in the computation of $\bm h_{u,v}^L$ in each step.
	\item As shown, the entities in each step are identical to the entities in the left bottom figure, implicitly encoding the subgraph representation.
\end{itemize}

\paragraph{Algorithm for path extraction.}

Given a drug pair $(u,v)$,
we use beam search to find the top $B=5$ paths in the direction from $u$ to $v$
and top $B=5$ paths from $v$ to $u$.
Take the direction from $u$ to $v$ as an example.
We provide the path extraction procedure in Algorithm~\ref{alg:path}.
We provide three kinds of lists:
openList, recording the top $K$ entities in each step;
closeList, recording the accumulated scores of entities visited in each step;
pathList, recording the searched paths at each step.
In lines~3-4,
we obtain the sets of entities visited in the $\ell$-th step $\mathcal V_{u,v}^{(\ell)}$
through bi-directional bread-first-search.
For each step,
we compute the accumulated scores of entities $e\in\mathcal V_{u,v}^\ell$
by summing the attention score $\alpha_r^{(\ell)}$ in lines~7-8,
and record the scores to the clostList.
Then we pick up edges with top-$B$ scores, and add them to openList and pathList
for next step computation in lines 11-13.
After $L$ steps,
we aggregate the selected paths in pathList[$1$], $\dots$, pathList[$L$]
to obtain the top-$B$ paths from $u$ to $v$.
The same steps are conducted to obtain the top-$B$ paths from $v$ to $u$.

\begin{algorithm}[H]
	\caption{Path extractor}
	\label{alg:path}
	\begin{algorithmic}[1]
		\REQUIRE {$(u,v), L, B$\qquad}\COMMENT{$B$: the number of top paths in each direction.}
		\STATE initialize openList$[0]\leftarrow u$;
		\STATE set $\mathcal V_{u,v}^{(0)}=\{u\}, \mathcal V_{u,v}^{(L)}=\{v\}$;
		\STATE obtain the set $\mathcal V_{u,v}^{(\ell)}=\{e: d(e,u)=\ell, d(e,v)=L-\ell\}, \ell=1, \dots, L$ with bread-first-search;
		\FOR{$\ell\leftarrow1$ to $L$}
		\STATE set closeList$[\ell]\leftarrow \emptyset$, pathList$[\ell]\leftarrow \emptyset$;
		\FOR{each edge in $\{(e',r,e): e'\in\text{openList}[\ell-1], e\in\mathcal V_{u,v}^{\ell} \}$}
		\STATE compute the attention weights $\alpha_{r}^{(\ell)}$;
		\STATE compute score($u,e', e$) = score($u,e$) + $\alpha_{r}^{(\ell)}$;
		\STATE closeList[$\ell$].add(($e$, score($u,e',e$)));
		\ENDFOR
		\FOR{$(u,e',e)\in$top$_B$(clostList[$\ell$])}
		\STATE openList[$\ell$].add($e$),  pathList[$\ell$].add($(e',r,e)$);
		\ENDFOR
		\ENDFOR
		\STATE \textbf{Return:} join(pathList[$1$]$\dots$pathList[$L$]).
	\end{algorithmic}
\end{algorithm}

\paragraph{Comparison of EmerGNN with other deep learning methods for link prediction.}

The general pipeline for GNN-based link prediction contains
three parts:
subgraph extraction,
node labeling,
and GNN learning.
Take SEAL~\citep{zhang2018link}
as an example.
It firstly extracts enclosing subgraph,
which contains the intersection of $L$-hop neighbors of $(u,v)$,
between drug pairs.
In order to distinguish the nodes on the subgraph,
SEAL then labels the nodes
based on their shortest path distance to both nodes $u$ and $v$.
Finally,
a GNN aggregates the representations of node labels for $L$ steps,
integrates the representations of all the nodes,
and predicts the interaction based on the integrated representation.

On subgraph extraction,
we use union of paths to form a path-based subgraph
instead of the enclosing subgraph,
as we need to integrate the entities and relations
on the augmented network 
while propagating from drug $u$ to $v$.
Node labeling is difficult to be extended to heterogeneous graph,
and a simple extension from homogeneous graph may not lead to good performance
\citep{teru2020inductive}.
In comparison,
our designed flow-based GNN avoids the labeling problem by propagating from $u$ to $v$
step-by-step.
Benefiting by the propagation manner,
we do not need use an extra pooling layer~~\citep{zhang2018link,vashishth2019composition,teru2020inductive}
and just use the pair-wise representation $\bm h_{u,v}^{(L)}$ in the last step
to encode  the path-based subgraph $\mathcal{G}_{u,v}^L$.
The above benefits are demonstrated in Fig.~5c
(main text) and we provide more analysis in Supplementary Figure~4.

\section{Additional Results}

\paragraph{Implementation of baselines.}

We summarize the details of how baseline methods are implemented for the DDI prediction tasks.

\begin{itemize}[leftmargin=*]
	\item \texttt{MLP} \citep{rogers2010extended}. 
	For each drug, there is a fingerprint vector with 1024 dimensions generated based on the drug's SMILES attributes,
	which stndw for Simplified Molecular Input Line Entry System.
	Given a pair of drugs $u$ and $v$,
	the fingerprints $\bm f_u$ and $\bm f_v$ are firstly fed into an MLP with 3 layers, respectively. 
	Then the representations are concatenated to compute the prediction logits with $\bm l(u,v) = \bm{W}_{\text{rel}}
	\big[\bm h_{u,v}^{(L)};\bm h_{v,u}^{(L)}\big]$.
	\item \texttt{Similarity} \citep{vilar2014similarity}.
	We generate four fingerprints based on the SMILES  representation for each drug. 
	For a given pair of drugs, we compute the similarity features between this drug pair and a known set of DDIs.
	Specifically, we compare the 16 pairwise similarity features composed of the fingerprints of each drug pair, 
	and select the maximum similarity value as the similarity feature for the current drug pair. 
	Subsequently, we input these features into a random forest model to predict the DDIs.
	
	\item \textcolor{black}{
		\texttt{CSMDDI} \citep{liu2022predict}.
		CSMDDI uses a RESCAL-based method to obtain embedding representations of drugs and DDI types. 
		It then utilizes partial least squares regression to learn a mapping function to bridge the drug attributes to their embeddings to predict DDIs.
		Finally, a random forest classifier is trained as the predictor, 
		and the output of the random forest classifier provides the final prediction score for the interaction between two drugs.
		The implementation follows  \url{https://github.com/itsosy/csmddi}.
	}
	
	\item \textcolor{black}{
		\texttt{STNN-DDI} \citep{yu2022stnn}.
		STNN-DDI learns a substructure×substructure×interaction tensor,
		which characterizes a substructure-substructure interaction (SSI) space, 
		expanded by a series of rank-one tensors. 
		According to a list of predefined substructures with PubChem fingerprint, 
		two given drugs are embedded into this SSI space. 
		A neural network is then constructed to discern the types of interactions triggered by the drugs 
		and the likelihood of triggering a particular type of interaction.
		The implementation follows \url{https://github.com/zsy-9/STNN-DDI}.
	}
	
	\item \texttt{HIN-DDI} \citep{tanvir2021predicting}.
	We constructs a heterogeneous information network (HIN) that integrates a biomedical network with DDIs.
	Within this network, we defined 48 distinct meta-paths, representing sequences of node types (including compounds, genes, and diseases) that connect nodes in the HIN. 
	For each meta-path, a series of topological features, such as path count, was generated. 
	Subsequently, these features were normalized and inputted into a random forest model for DDI prediction.
	
	\item \texttt{MSTE} \citep{yao2022effective}.
	MSTE learns DDI with knowledge graph embedding technique
	and models the interactions as triplets in the KG.
	Specifically,
	for each interaction $(u,i,v)\in\mathcal N_\text{D}$,
	there are learnable embedding vectors 
	$\bm e_u, \bm e_v\in\mathbb R^d$ for the drugs $u$ and $v$, respectively,
	and $\bm i\in\mathbb R^d$ for interaction type $i$.
	MSTE then computes a score
	$s(u,i,v) = \|\sin(\bm i\cdot\bm e_v)\cdot\bm e_u + 
	\sin(\bm e_u\cdot \bm e_v)\cdot \bm i - \sin(\bm e_u\cdot \bm i)\cdot \bm e_v  \|_{1/2}$,
	which is then used as a negative logit for the prediction of 
	interaction type $i$.
	The dimension $d$ is a hyper-parameter tuned among \{32, 64, 128\}.
	The implementation follows \url{https://github.com/galaxysunwen/MSTE-master}.
	
	\item \texttt{KG-DDI} \citep{karim2019drug}.
	KG-DDI uses a Conv-LSTM network on top of the embeddings
	to compute the score of interaction triplets $(u,i,v)\in\mathcal N_\text{D}$ as well as the biomedical triplets $(h,r,t)\in\mathcal N_\text{B}$.
	Different from MSTE,
	KG-DDI firstly optimizes the parameters on both the interaction triplets
	and biomedical triplets,
	namely triplets in the augmented network,
	then fine-tunes on the interaction triplets for final prediction.
	The implementation follows
	\url{https://github.com/rezacsedu/Drug-Drug-Interaction-Prediction}.
	
	\item \texttt{CompGCN} \citep{vashishth2019composition}.
	All the drugs, biomedical concepts, interactions and relations
	have their own learnable embeddings.
	These embeddings are aggregated by a graph neural network with 1 layer.
	The high-order embeddings $\bm h_u^L, \bm h_v^L, \bm h_i^L$
	are used to compute the score $s(u,i,v) =\langle\bm h_u^L, \bm h_v^L, \bm h_i^L\rangle$, which is then used as the logic of interaction type $i$.
	The implementation follows \url{https://github.com/malllabiisc/CompGCN}.
	
	\item \texttt{Decagon} \citep{zitnik2018modeling}.
	Decagon is similar to CompGCN.
	The main difference is that the input biomedical network
	only considers biomedical concepts of drugs, genes and diseases,
	rather than the full biomedical network $\mathcal N_\text{B}$.
	
	\item \texttt{KGNN} \citep{lin2020kgnn}.
	KGNN is built upon a GNN which propagates information 
	and learns node representations within the new knowledge graph. Considering computational efficiency, KGNN employed neighbor sampling, with four neighbors sampled per layer for a total of two layers. Subsequently, the learned node representations were used to predict DDIs. The implementation follows \url{https://github.com/xzenglab/KGNN}.
	
	\item \texttt{SumGNN} \citep{yu2021sumgnn}.
	SumGNN has three steps.
	First,
	we extract enclosing subgraphs from the augmented network
	for all the drug pairs $(u,v)$
	to be predicted.
	Second,
	a node labeling trick is applied for all the enclosing subgraphs
	to compute the node features.
	Then, a graph neural network computes the graph representations of enclosing subgraphs, which are finally used to predict the interaction.
	The implementation follows \url{https://github.com/yueyu1030/SumGNN}.
	
	\item 
		\texttt{DeepLGF} \citep{ren2022biomedical}.
		The DeepLGF model contains three parts. 
		First, the SMILES of drugs are used as sentences to encode the drugs' chemical structure.
		Second, a KG embedding model ComplEx is applied on the biomedical network to get the global embedding information of drugs.
		Third, a relational-GNN is used to aggregate the representations from the biomedical network.
		Finally, the three kinds of representations are fused with an MLP module for the DDI prediction.
		Since there is no official code provided, we implement this model based on CompGCN.
	
\end{itemize}

\paragraph{Performance comparison of the S0 setting.}

There are three basic settings for the DDI prediction
\citep{dewulf2021cold,liu2022predict,yu2022stnn}:
(S0) interaction between existing  drugs;
(S1) interaction between emerging and existing drugs; and
(S2) interaction between emerging  drugs.

\texttt{EmerGNN} has shown substantial advantage over the baseline methods for emerging drug prediction in  Table~1 (main text).
We also compare the performance in the S0 setting
for prediction between existing drugs in Supplementary Table~\ref{tab:existing},
where the setting exactly follows \cite{yu2021sumgnn}.
Comparing the two tables,
we find that the emerging drug prediction task is much harder than existing drug prediction
as the accuracy values in the Table~1 (main text) are much lower than those in Supplementary Table~3.
Even though the shallow models \texttt{MLP}, \texttt{Similarity}, \texttt{HIN-DDI}
perform well in predicting DDIs for emerging drugs,
they are worse than the deep networks when predicting DDIs between existing drugs.
The embedding model \texttt{MSTE} performs very poorly for emerging drugs
but is the third best for existing drug prediction.
The GNN-based methods, especially \texttt{SumGNN},
also works well for predicting DDIs between existing drugs.
This demonstrates that drug embeddings and deep networks
can be helpful for drug interaction prediction
if sufficient data are provided.
\texttt{EmerGNN}, even though specially designed for emerging drug prediction,
still outperforms the baselines with a large margin
for predicting interactions between existing drugs.
These results again show the flexibility and strengths of \texttt{EmerGNN}
on the DDI prediction task.

\paragraph{Path visualization.}
\label{app:visualize}


We provide additional results for path visualization
between the case of S1 setting and S0 setting (Supplementary Figure~\ref{fig:app-visualization}).
Specifically,
we choose examples with predicted interaction types
\#52, \#5 and \#18 in Supplementary Table~\ref{tab:correlation}.
We plot the interactions between emerging and existing drugs in the left part,
and the interactions between two existing drugs in the right part.
As shown in Supplementary Figure~\ref{fig:app-visualization}, relation type CrC plays an important role during prediction,
which is also reflected by the high correlations in Fig.~3c (main text).
For the interaction types on the subgraphs,
we also observe the correlations of interaction types,
namely (\#52, \#39), (\#5, \#85) and (\#18, \#49),
which are identified in Supplementary Table~\ref{tab:correlation}.
Comparing the left part with right part,
we observe that the biomedical entities,
like Gene::1565, Disease::DOID:10763 and Parmacologic Class::N000000102(9),
play the role to connect the emerging drug and existing drug.
However,
the prediction of interactions between two existing drugs relies mainly on 
the DDI between drugs.
These results again verify the claim
that EmerGNN is able to identify and leverage the relevant entities and relations in the biomedical network.

\section*{Supplementary Tables}

\begin{table}[H]
	\centering
	\color{black}
	\caption{Statistics of datasets used for predicting interactions for DDI prediction. 
		$\mathcal V$'s represent the sets of nodes.
		$\mathcal R$'s represent the sets of interaction types.
		$\mathcal N$'s represent the sets of edges.
	}
	\small
	\vspace{3px}
	
	\leftline{\qquad\qquad\textbf{S0 setting:} prediction interactions between exiting drugs.}
	\label{tab:SI:stats}
	\begin{tabular}{c|c|c|ccc}
		\toprule
		Statistics & $\mid\!\!\mathcal V_\text{D}\!\!\mid$ &  $\mid\!\!\mathcal R_\text{I}\!\!\mid$ & $\mid\!\!\mathcal N_\text{D-train}\!\!\mid$ & $\mid\!\!\mathcal N_\text{D-valid}\!\!\mid$ & $\mid\!\!\mathcal N_\text{D-test}\!\!\mid$  \\ 	\midrule
		Drugbank   &  1,710  &  86 & 134,641  & 19,224 &  38,419   \\
		TWOSIDES   & 604  &  200 & 177,568 & 24,887 & 49,656  \\
		\bottomrule       
	\end{tabular}
	\vspace{8px}
	
	\setlength\tabcolsep{2.5pt}
	\leftline{\textbf{S1 and S2 settings:} predicting interactions for emerging drugs.}
	\resizebox{\textwidth}{!}{
		\begin{tabular}{cl|ccc|c|c|cc|cc}
			\toprule
			\multirow{2}{*}{Data}  &  \multirow{2}{*}{seed}  &  \multirow{2}{*}{$\mid\!\!\mathcal V_\text{D-train}\!\!\mid$} &  \multirow{2}{*}{$\mid\!\!\mathcal V_\text{D-valid}\!\!\mid$} & \multirow{2}{*}{$\mid\!\!\mathcal V_\text{D-test}\!\!\mid$} &
			\multirow{2}{*}{$\mid\!\!\mathcal R_\text{I}\!\!\mid$} & 
			\multirow{2}{*}{$\mid\!\!\mathcal N_\text{D-train}\!\!\mid$}  &  
			\multicolumn{2}{c|}{S1}  & \multicolumn{2}{c}{S2} \\
			&&&&&&
			&$\mid\!\!\mathcal N_\text{D-valid}\!\!\mid$ &  $\mid\!\!\mathcal N_\text{D-test}\!\!\mid$ 
			&   $\mid\!\!\mathcal N_\text{D-valid}\!\!\mid$ &  $\mid\!\!\mathcal N_\text{D-test}\!\!\mid$ 
			\\ 	\midrule
			\multirow{5}{*}{DrugBank} 
			&	1 	 &  1,461   &  79   & 161   &    86  &		137,864   &   17,591   &   32,322     &  536 &  1,901  \\
			&	 12	 &  1,465   &  79   &  161  & 86   & 140,085   &    17,403    &   30,731     & 522 &  1,609  \\
			&	123 	 &  1,466   &  81   &  161  &   86 &  140,353    &   14,933     &   32,845   & 396 &   1,964  \\
			&	 1234	 & 1,463    & 81    & 162   & 86   & 139,141     &  15,635  &  33,254  &  434 & 1,956   \\
			&	 12345	 &   1,461  &  80   &  169  &  86  & 133,394     &   17,784     &  35,803     & 546  &   2,355 \\ 		\midrule
			\multirow{5}{*}{TWOSIDES}   &	1 	 &  514   &   30  &  60 & 200 &  185,673  &   16,113   &   45,365     & 467 & 2,466      \\
			&	12 	 &  514   &   30  &  60  &  200	&  172,351   &	23,815	&  48,638  & 717 & 3,373    \\
			&	123    &  514   & 30    &  60  &  200  &   181,257   &   18,209    &   46,969   &  358 &  2,977   \\
			&	1234	 & 514    &  30   & 60   & 200  &  186,104	&  25,830  &  35,302   &  837 & 1,605   \\
			&	12345 	 &  514   &  30   &  60  &  200  &   179,993   &   22,059     &   43,867    & 702 & 2,695   \\
			\bottomrule       
		\end{tabular}
	}
	\vspace{8px}
	
	\setlength\tabcolsep{6pt}
	\leftline{\textbf{Biomedical network:} HetioNet~\cite{himmelstein2017systematic} is used in this paper.}
	\resizebox{\textwidth}{!}{
		\begin{tabular}{cl|cccccc}
			\toprule
			Data & Seed  & $\mid\!\!\mathcal V_\text{B}\!\!\mid$	&	$\mid\!\!\mathcal R_\text{B}\!\!\mid$	 & $\mid\!\!\mathcal N_\text{B}\!\!\mid$ &$\mid\!\!\mathcal N_\text{B-train}\!\!\mid$  & $\mid\!\!\mathcal N_\text{B-valid}\!\!\mid$ & $\mid\!\!\mathcal N_\text{B-test}\!\!\mid$ \\ 	\midrule
			\multirow{5}{*}{DrugBank}&  1 &  34,124 &  23 & 1,690,693  &1,656,037   & 1,666,317 & 1,690,693 \\
			&  12 &  34,124 &  23 & 1,690,693  &1,658,075   & 1,668,273 & 1,690,693 \\
			&  123 &  34,124 &  23 & 1,690,693  &1,657,489   & 1,667,685 & 1,690,693 \\
			&  1234 &  34,124 &  23 & 1,690,693  &1,657,400   & 1,668,685 & 1,690,693 \\
			&  12345 &  34,124 &  23 & 1,690,693  &1,656,603   & 1,668,091 & 1,690,693 \\
			\midrule
			\multirow{5}{*}{TWOSIDES}&  1 &  34,124 &  23 & 1,690,693  &1,671,519   & 1,678,548 & 1,690,693 \\
			&  12 &  34,124 &  23 & 1,690,693  &1,669,693   & 1,676,696 & 1,690,693 \\
			&  123 &  34,124 &  23 & 1,690,693  &1,672,632   & 1,678,335 & 1,690,693 \\
			&  1234 &  34,124 &  23 & 1,690,693  &1,671,617   & 1,678,528 & 1,690,693 \\
			&  12345 &  34,124 &  23 & 1,690,693  &1,672,288   & 1,678,776 & 1,690,693 \\
			\bottomrule       
		\end{tabular}
	}
\end{table}

\begin{table}[H]
	\centering
	\caption{Hyper-parameters and their tuning ranges for hyper-parameter selection.
		For all the baselines and the proposed EmerGNN,
		we use the hyper-parameter optimization toolbox hyperopt \citep{bergstra2015hyperopt}
		to search for the optimum among 360 of hyper-parameter configurations.
		The objective of hyper-parameter selection is to 
		maximize the premier metric performance (F1-score in DrugBank and PR-AUC in TWOSIDES)
		on the validation data.
		Adam \citep{kingma2014adam} is used as the optimizer to update the model parameters of EmerGNN.}
	\small
	\label{tab:hyperparameter}
		\begin{tabular}{c|c}
			\toprule
			Hyper-parameter & Ranges \\
			\midrule
			Learning rate & $\{1\times 10^{-4}, 3\times 10^{-4}, 1\times10^{-3},  3\times 10^{-3}, 1\times 10^{-2}\}$\\
			Weight decay rate &  $\{1\times10^{-8}, 1\times10^{-6}, 1\times10^{-4}, 1\times10^{-2}\}$  \\
			Mini-batch size &  $\{32, 64, 128\}$ \\
			Representation size $d$ & $\{32, 64\}$ \\
			Length of subgraphs $L$ & $\{2,3,4\}$ \\
			\bottomrule
		\end{tabular}
\end{table}

\begin{table}[H]
	\centering
	\caption{Comparison of different methods on the DDI prediction between two existing drugs (S0 Setting). 
		``DF'' is short for ``Drug Feature'';
		``GF'' is short for ``Graph Feature'';
		``Emb'' is short for ``Embedding''; and
		``GNN'' is short for ``Graph Neural Network''.}\tablefootnote[1]{All of the methods are run for five times on the five-fold datasets
		with mean value and standard deviation reported on the testing data.
		The evaluation metrics are presented in percentage (\%)
		with the larger value indicating better performance.
		The boldface numbers indicate the best values,
		while the underlined numbers indicate the second best.
		p-values are computed under two-sided t-testing of 
		EmerGNN over the second best baselines.}
	\small
	\setlength\tabcolsep{4pt}
	\renewcommand\arraystretch{1}
	\label{tab:existing}
	\resizebox{\textwidth}{!}{
		\begin{tabular}{ll|ccc|ccc}
			\toprule
			\multicolumn{2}{c|}{Datasets (S0 setting)}               &                        \multicolumn{3}{c|}{\textbf{DrugBank}}                        &                        \multicolumn{3}{c}{\textbf{TWOSIDES}}                         \\ \midrule
			Type & Methods               &          F1-Score          &          Accuracy          &      Kappa           &           PR-AUC       &          ROC-AUC            &           Accuracy            \\ \midrule
			DF	&	MLP \citep{rogers2010extended}   &       61.1$\pm$0.4       &       82.1$\pm$0.3       &       80.5$\pm$0.2        &       81.2$\pm$0.1    &       82.6$\pm$0.3          &       73.5$\pm$0.3       \\
			&	Similarity \citep{vilar2014similarity} & 55.0$\pm$0.3 & 62.8$\pm$0.1  & 67.6$\pm$0.1 & 59.5$\pm$0.0   & 59.8$\pm$0.0  &  57.0$\pm$0.1 \\ \midrule
			GF	&	HIN-DDI \citep{tanvir2021predicting}	&  46.1$\pm$0.5 &  54.4$\pm$0.1 &  63.4$\pm$0.1 &  83.5$\pm$0.2 &  87.7$\pm$0.3  &  82.4$\pm$0.3  \\  \midrule
			Emb	&	MSTE \citep{yao2022effective} 	&  83.0$\pm$1.3 &  85.4$\pm$0.7 & 82.8$\pm$0.8  &	  90.2$\pm$0.1	&	91.3$\pm$0.1	&	84.1$\pm$0.1   \\ 
			&	KG-DDI \citep{karim2019drug}    &       52.2$\pm$1.1       &       61.5$\pm$2.8       &       55.9$\pm$2.8         &       88.2$\pm$0.1      &       90.7$\pm$0.1      &       83.5$\pm$0.1       \\ \midrule
			GNN	& 	CompGCN \citep{vashishth2019composition}      &  74.3$\pm$1.2   &  78.8$\pm$0.9  &	 75.0$\pm$1.1   &  90.6$\pm$0.3 &   92.3$\pm$0.3 & 84.8$\pm$0.3\\
			&	Decagon	\citep{zitnik2018modeling} &       57.4$\pm$0.3       &       87.2$\pm$0.3       &       86.1$\pm$0.1         &       90.6$\pm$0.1   &       91.7$\pm$0.1          &       82.1$\pm$0.5       \\
			&	KGNN \citep{lin2020kgnn}      &       74.0$\pm$0.1       &       90.9$\pm$0.2       &       89.6$\pm$0.2       &       90.8$\pm$0.2   &       92.8$\pm$0.1           &       86.1$\pm$0.1       \\
			&	SumGNN \citep{yu2021sumgnn}     & \underline{86.9$\pm$0.4} & \underline{92.7$\pm$0.1} & \underline{90.7$\pm$0.1} & \underline{93.4$\pm$0.1}	& \underline{94.9$\pm$0.2}  & \underline{88.8$\pm$0.2} \\ 
			& {\textbf{EmerGNN}}            &  \textbf{94.4}$\pm$0.7   &  \textbf{97.5}$\pm$0.1   &  \textbf{96.6}$\pm$0.8  &  \textbf{97.6}$\pm$0.1    &  \textbf{98.1}$\pm$0.1   &  \textbf{93.8}$\pm$0.2   \\ \midrule
			\multicolumn{2}{l|}{\textbf{p-values}} 	&           4.5E-7            &           6.5E-13            &           6.7E-8          &           2.3E-8     &           5.1E-10                     &           6.1E-7            \\ \bottomrule
		\end{tabular}
	}
\end{table}

\begin{table}[H]
	\caption{Complexity analysis of different GNN-based methods in the S1 setting in terms of GPU memory footprint and the number of model parameters.}
	\resizebox{\textwidth}{!}{
		\begin{tabular}{c|cc|cc}
			\toprule
			&\multicolumn{2}{c|}{\textbf{DrugBank}}  & \multicolumn{2}{c}{\textbf{TWOSIDES}} \\
			&  GPU memory (MB)  & Model parameters & GPU memory (MB) & Model parameters \\
			\midrule
			Decagon &  8,214  & 1,766,492  & 2,908  & 1,145,850 \\
			SumGNN & 6,968   & 1,237,628  & 12,752 & 1,263,188  \\
			DeepLGF &  16,822  & 11,160,226  & 5,974  & 10,012,456 \\
			EmerGNN & 7,104   & 137,164  & 2,040 & 156,406  \\
			\bottomrule
		\end{tabular}
	}
\end{table}

\begin{table}[H]
	\centering
	\caption{Detailed explanation on the highlighted interaction pairs (three yellow circles) in Fig.~3a (main text).
		The interaction IDs are from the original DrugBank datasets.}
	\label{tab:correlation}
	\small
	\setlength\tabcolsep{3pt}
	\renewcommand\arraystretch{1.2}
	\resizebox{\textwidth}{!}{
		\begin{tabular}{c|L{260px}|l}
			\hline
			ID & Description                                  &           Exemplar drug-pairs                 \\ \hline
			\#52 &  \#Drug1 may decrease the analgesic activities of \#Drug2. &   (Tapentadol, Dolasetron)        
			\\ 	\hline
			\#39 & \#Drug1 may increase the constipating activities of \#Drug2. &     (Cyclopentolate, Ramosetron)                   \\ \hline 
			Evidence & \multicolumn{2}{L{410px}}{Oral naloxone is efficacious in reversing opioid-induced constipation, but often causes the unwanted side effect of analgesia reversal \citep{liu2002low}.}  \\ \hline
			\hline
			\#5 & \#Drug1 may decrease the vasoconstricting activities of \#Drug2.  &        (Labetalol, Formoterol)                
			\\ 	\hline
			\#85 & \#Drug1 may increase the tachycardic activities of \#Drug2. &          (Duloxetine, Droxidopa)              \\ \hline 
			Evidence & \multicolumn{2}{L{410px}}{This decrease in afferent signaling from the baroreceptor causes  vasoconstriction and increased heart rate (tachycardic) \citep{chaudhry2022physiology}.} \\ 
			\hline
			\hline
			\#18 & \#Drug1 can cause an increase in the absorption of \#Drug2 resulting in an increased serum concentration and potentially a worsening of adverse effects.	& (Ethanol, Levomilnacipran)     \\		\hline
			\#49 & The risk or severity of adverse effects can be increased when \#Drug1 is combined with \#Drug2.& (Methyl salicylate, Triamcinolone)  	\\ \hline
			Evidence & \multicolumn{2}{L{410px}}{Both of the two interactions are related to worsening adverse effects when two drugs are combined together.} \\ 
			\hline  
		\end{tabular}
	}
\end{table}

\begin{table}[H]
	\centering
	\caption{Detailed explanation of selected paths from the learned model.
		In these cases, we provide the target interaction sample to be predicted, two important paths selected by our method, and the corresponding explanations.}
	\small
	
	\leftline{\textbf{Case 1 in Fig.~4a.}}
	\begin{tabular}{L{\textwidth}}
		\toprule
		\textbf{Target}: Tapentadol (DB06204) may decrease the analgesic activity of Dolasetron (DB00757).
		\\
		\midrule
		\textbf{Path1} (0.6666): Tapentadol$\xrightarrow{\text{binds}}$CYP2D6 (P450)$\xrightarrow{\text{binds\_inv}}$Dolasetron\\
		\hline
		Explanation: Tapentadol can binds the P450 enzyme CYP2D6 (Gene::1565), which is vital for the metabolism of  many drugs like Dolasetron (Estabrook, 2003). In addition, Binding of drug to plasma proteins is reversible, and changes in the ratio of bound to unbound drug may lead to drug-drug interactions (Kneip et. al. 2008).
		\\
		\midrule
		\textbf{Path2} (0.8977): Dolasetron$\xrightarrow{\text{resembles}}$ Hyoscyamine$\xrightarrow{\text{\#39:}\uparrow\text{constipating}}$Eluxadoline $\xrightarrow{\text{\#39\_inv}}$Tapentadol
		\\
		\hline
		Explanation: Dolasetron is similar to drug Hyoscyamine (DB00424). Hyoscyamine and Tapentadol can get some connection since they will both increase the constipating activity of Eluxadoline (DB09272). As suggested by Liu and Wittbrodt (2022), reversing opioid-induced constipation often causes the unwanted side effect of analgesia reversal.
		\\
		\bottomrule
	\end{tabular}
	\vspace{8px}
	
	\leftline{\textbf{Case 2 in Fig.~4a.}}
	\begin{tabular}{L{\textwidth}}
		\toprule
		\textbf{Target}: Labetalol (DB00598) may decrease the vasoconstricting activity of Metaraminol (DB00610).
		\\
		\midrule
		\textbf{Path1} (0.8274): Labetalol$\xrightarrow{\text{resembles}}$Isoxsuprine$\xrightarrow{\text{\#8\_inv}}$Dronabinol $\xrightarrow{\text{\#85:}\uparrow\text{tachycardic}}$Metaraminol
		\\
		\hline
		Explanation: Lebetalol is similar to the drug Isoxsupirune (DB08941). Isoxsupirune and Metaraminol can get some connection since Dronabinol (DB00470) will increase the tachycardic activity of both of them. As suggested by Chaudhry et al (2022), the decrease of vasoconstriction and the increase of tachycardic are often correlated. \\
		\midrule
		\textbf{Path2} (0.8175): Metaraminol$\xrightarrow{\text{\#5:}\downarrow\text{vasoconstricting\_inv}}$Spironolactone$\xrightarrow{\text{treat}}$hypertension $\xrightarrow{\text{treat\_inv}}$Labetalol  \\ \hline
		Explanation: Labetelol and Spironolactone (DB00421) get can some  connections since they treat the same disease hypertension (DOID:10763). As Spironolactone may decrease the vasoconstricting activity of Metaraminol (indicated by the inverse edge), we predict that Labetelol may also decrease the vasoconstricting activity of Metaraminol.
		\\
		\bottomrule
	\end{tabular}
\end{table}

\begin{table}[H]
	\centering
	\caption{Performance comparison of different technique designing in EmerGNN on DrugBank dataset.
		``Undirected edges w.o. inverse'' means the variant that uses undirected edges instead of introducing the inverse edges.
		``Subgraph representation'' means the variant that learns a subgraph representation as \texttt{SumGNN} upon $\mathcal{G}_{u,v}^L$.
		``Uni-directional pair-wise representation'' means the variant that only learns on the uni-directional computing (Method) from direction $u$ to $v$ without considering the direction from $v$ to $u$.
		The performance of EmerGNN is provided in the last row as a reference.
	}
	\begin{tabular}{c|ccc}
		\hline
		Variants of designing  & F1-Score              & Accuracy             & Kappa        \\		\hline
		Undirected edges w.o. inverse  & 53.7$\pm$2.0  & 61.8$\pm$1.9  &  54.8$\pm$2.0 \\
		Subgraph representation  &    33.1$\pm$3.6  &     50.2$\pm$5.6    & 	40.7$\pm$5.6	 \\ 
		Uni-directional	pair-wise representation	&   55.6$\pm$2.1   &   67.4$\pm$1.6  	 &  61.1$\pm$1.6  		 \\
		\hline
		EmerGNN  &  {62.0}$\pm$2.0    &    {68.6}$\pm$3.7 	 &    	{62.4}$\pm$4.3	 \\
		\hline
	\end{tabular}
\end{table}

\clearpage
\section*{Supplementary Figures}

\begin{figure}[H]
	\centering
	\includegraphics[width=\textwidth]{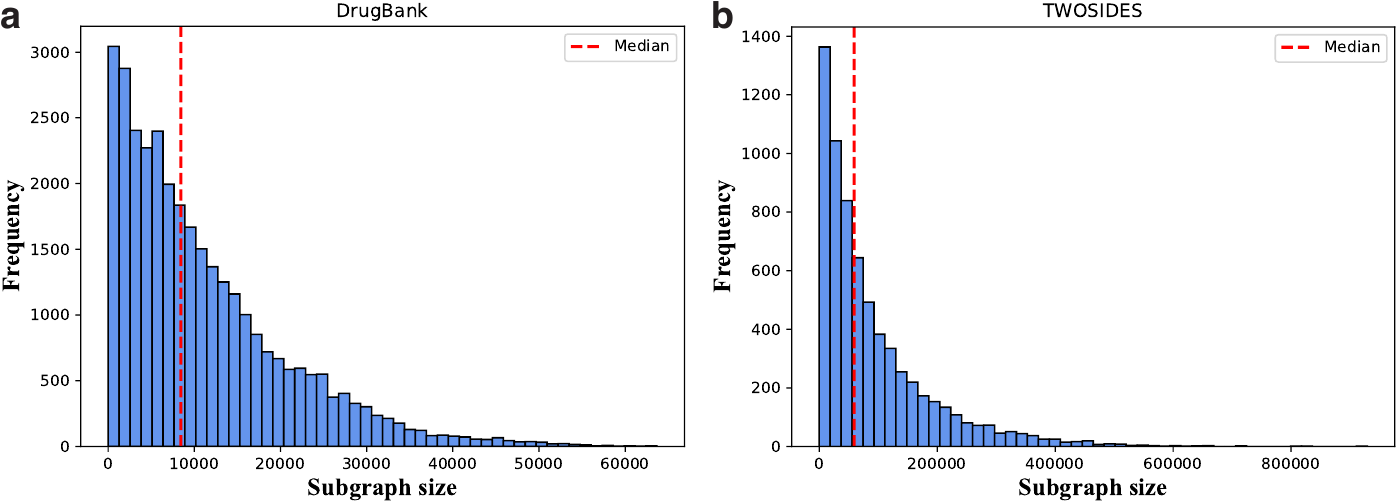}
	\caption{Histograms of subgraph sizes of $\mathcal G_{u,v}^L$ (indicated by the number of edges) in the testing sets of two datasets when $L=3$. 
		Median values (8,444 for DrugBank and 59120 for TWOSIDES) are indicated by the red dashed line.}
	\label{fig:gsize}
\end{figure}

\begin{figure}[H]
	\centering
	\color{black}
	\includegraphics[width=\textwidth]{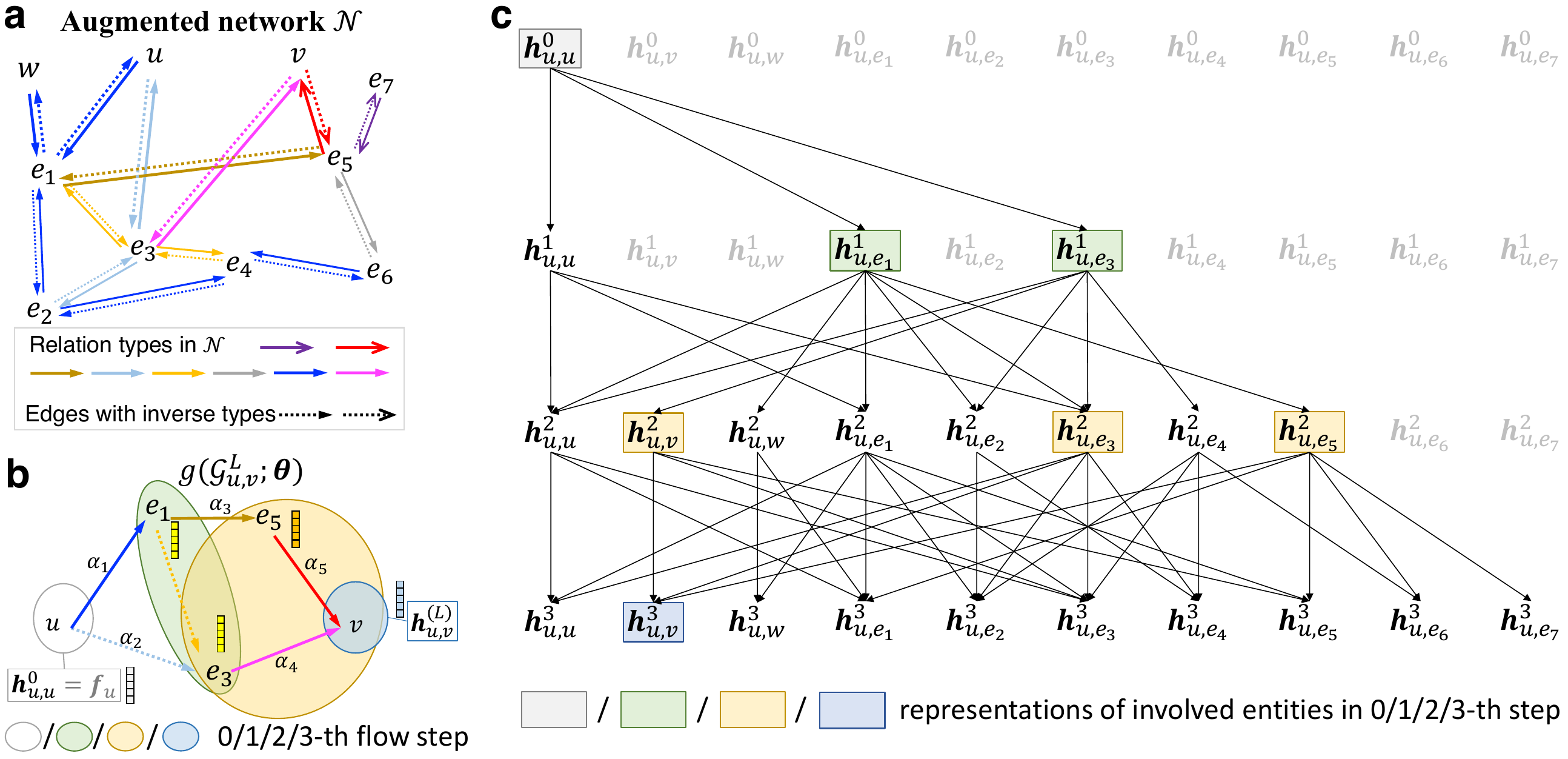}
	\caption{A graphical illustration of why the 
		initialization step together with the
		message propagation function can implicitly encode the visible entities in each layer (step $\ell$).
		\textbf{(a)} Symbolic representation of the augmented network in the example of Fig.~1 (main text). Different colors in edges mean different relation types (in Fig.~1a) and the dashed lines mean the inverse edges with corresponding relation type.
		\textbf{(b)} Symbolic representation of the flow-based GNN from $u$ to $v$. The four circles in different colors indicate the involved entities in different steps.
		\textbf{(c)} Representation flows according to the proposed Algorithm~1 (gray symbols mean $\bm 0$ vectors, and the relation types in lines are omitted for simplicity).
		From top to bottom, we show how the representations are activated and propagated in each step.
		The involved entities in each step in (b) and (c) are identical to each other,
		indicating that our algorithm can implicitly encoding the subgraph representation.
	}
	\label{fig:illustration}
\end{figure}

\begin{figure}[H]
	\centering
	\includegraphics[width=0.6\textwidth]{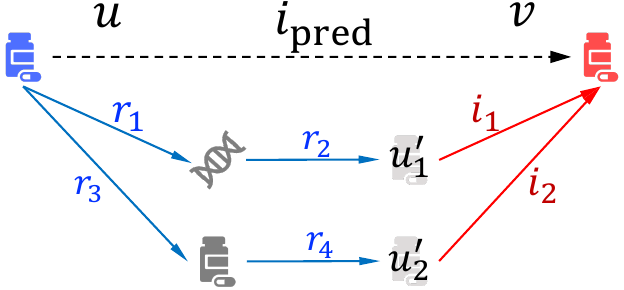}
	\caption{A graphical exampler of selected paths (the different icons mean different drugs and genes).
		We show how the correlation matrices in Fig.~3a and 3c (main text) are calculated based on this example.
		Given the interaction triplet $(u,i_\text{pred},v)$ to be predicted,
		we extract several paths (two in this figure) through Algorithm~\ref{alg:path}.
		Here, we have two paths which contain some relations
		\{$r_1,\dots,r_4$\} in the biomedical network $\mathcal N_B$
		and interactions \{$i_1, i_2$\} in the interaction network $\mathcal N_D$.
		The co-occurrence times for each type $i\in\mathcal R_\text{I}$  and $r\in\mathcal R_\text{B}$
		are counted on the paths for different interaction triplets.
		For the interaction types $i\in\mathcal R_\text{I}$ or biomedical relation types $r\in\mathcal R_\text{B}$,
		we group their counting values according to the to-be-predicted interaction $i_\text{pred}$
		and normalize the values by dividing the frequency of $i_\text{pred}$ in $\mathcal N_\text{D-test}$.}
	\label{fig:path-exmp}
\end{figure}

\begin{figure}[H]
	\centering
	\includegraphics[width=\textwidth]{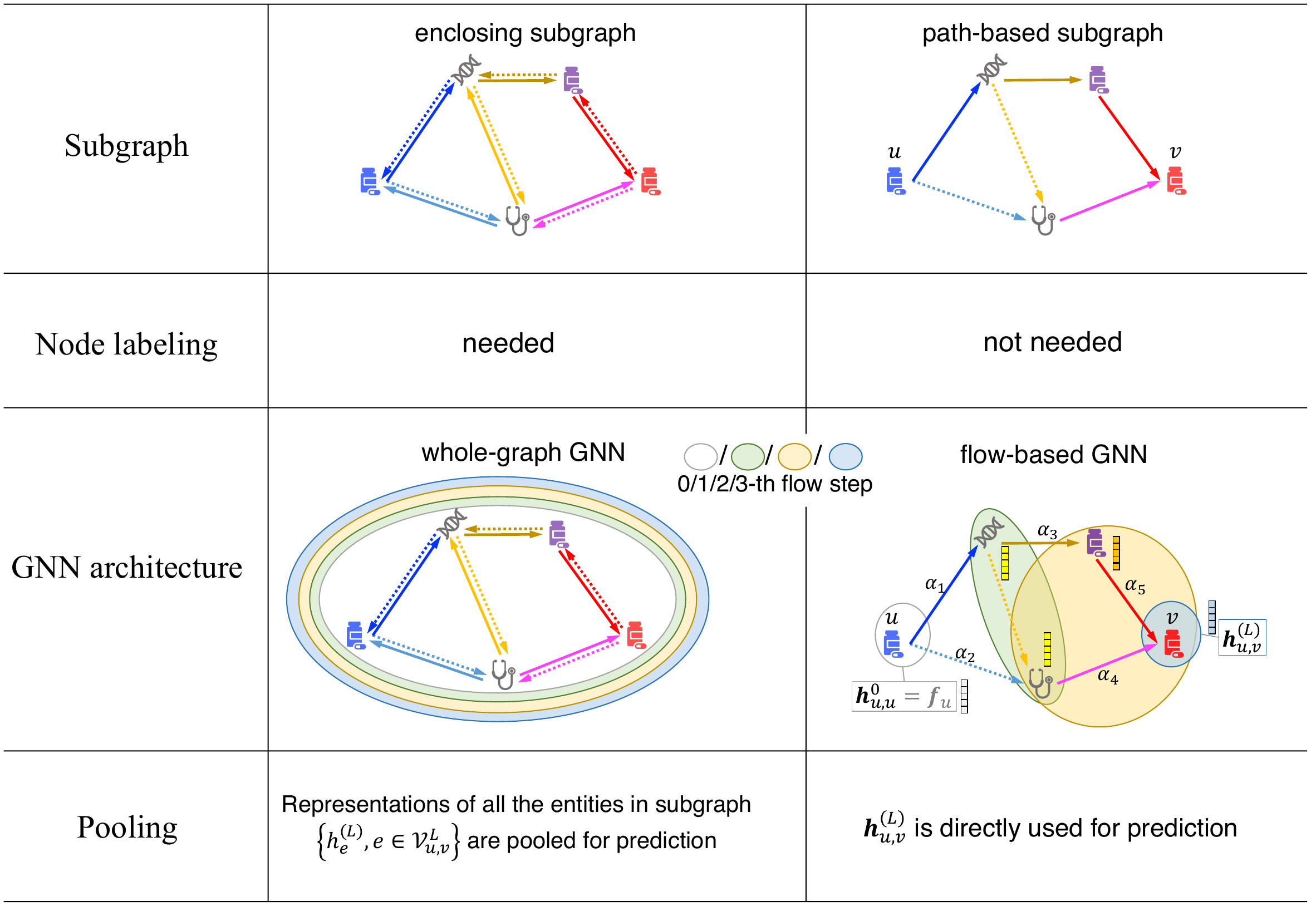}
	\caption{A detailed comparison between SumGNN and EmerGNN in terms of subgraph structure, usage of node labeling, GNN architecture design and the pooling mechanism.
		Different colors in the edges indicate different relation types.
		The different circles mean different computing steps in GNNs.
		\textbf{Subgraph}: The enclosing subgraph used in SumGNN contains all the edges among entities within $L$ steps away from both $u$ and $v$;
		the path-based subgraph only considers edges pointing from $u$ to $v$ or $v$ to $u$.
		\textbf{Node labeling}: SumGNN requires a node labeling procedure to compute the distance of nodes on the subgraph to the target drugs $u$ and $v$;
		however, as the edges are connected in the direction from $u$ to $v$, in EmerGNN, the distance information can be reflected by the number of jumps, 
		thus EmerGNN does not need a node labeling procedure.
		\textbf{GNN architecture}: SumGNN uses the whole-graph GNN as in \citep{zhang2018link,xu2018powerful,teru2020inductive} to propagate over the whole subgraph. 
		EmerGNN uses the flow-based GNN to propagate information from $u$ to $v$ step-by-step
		with a better control of information flow.
		\textbf{Pooling}: In SumGNN, the representations of all the entities in the subgraph
		should be pooled (for example concatenated) for final interaction prediction;
		however, benefiting from the flowing pattern of flow-based GNN, all the information can be ordered and integrated when propagating from $u$ to $v$,
		thus EmerGNN only uses the final step representation of $v$, namely
		$\bm h_{u,v}^{(L)}$, for interaction prediction.
	}
\end{figure}

\begin{figure}[H]
	\centering
	\includegraphics[width=\textwidth]{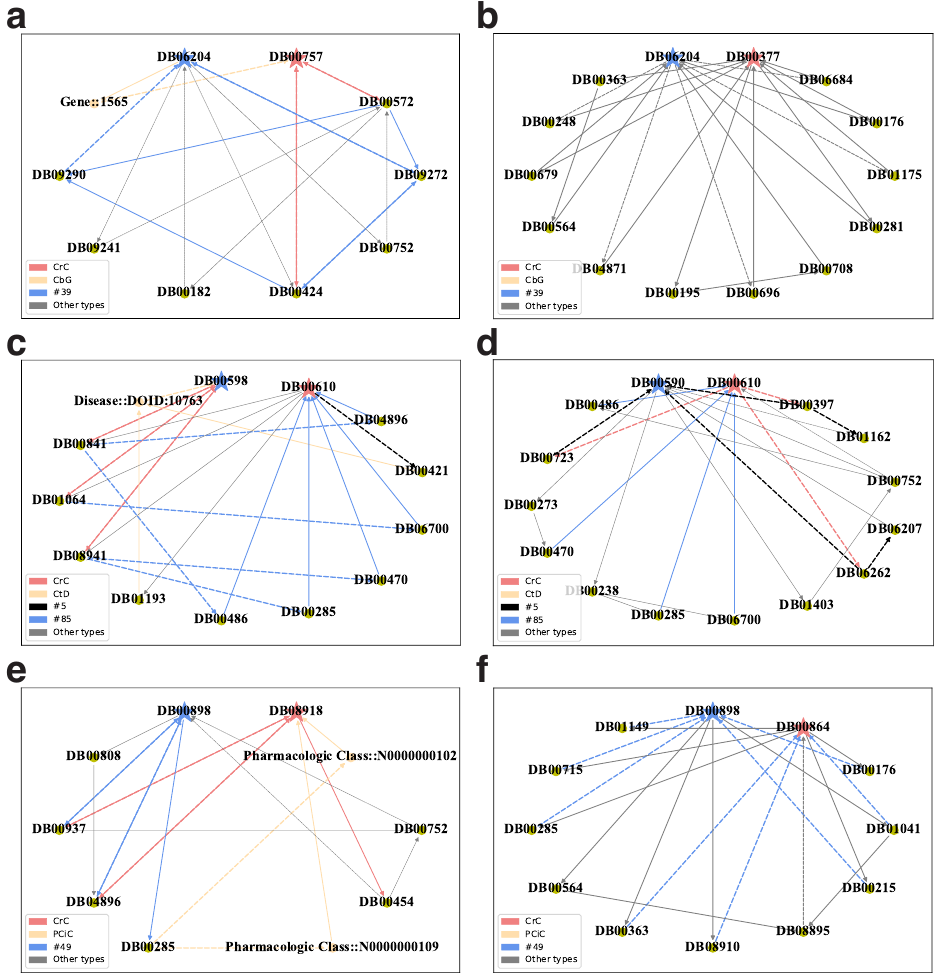}	
	\caption{Visualization of learned paths between drug pairs.
		Left: interactions between an existing drug and an emerging drug.
		Right: interactions between two existing drugs.
		The dashed lines mean inverse types.
		\textbf{(a,b)} DB00757 (Dolasetron) is an emerging drug.
		DB06204 (Tapentadol) and DB00377 (Palonosetron) are existing drugs.
		The prediction interaction type is 
		\#52 (\#Drug1 may decrease the analgesic activities of \#Drug2.).
		\textbf{(c,d)} DB00598 (Labetalol) is an emerging drug.
		DB00610 (Metaraminol) and DB00590 (Doxazosin) are existing drugs.
		The prediction interaction type is 
		\#5 (\#Drug1 may decrease the vasoconstricting activities of \#Drug2.).
		\textbf{(e,f)}  DB08918 (Levomilnacipran) is an emerging drug.
		DB00898 (Ethanol) and DB00864 (Tacrolimus) are existing drugs.
		The prediction interaction type is 
		\#18 (\#Drug1 can cause an increase in the absorption of \#Drug2 resulting in an increased serum concentration and potentially a worsening of adverse effects.).
	}
	\label{fig:app-visualization}
\end{figure}

\end{document}